\documentstyle[12pt,fleqn]{article}

\begin{document}     

\baselineskip= 8mm


\begin{center}
{\Large\bf Finite temperature nonlocal effective action\\[5mm]  
for  quantum  fields in  curved  space}
\end{center}
\vspace{6mm}          
\begin{flushright}
Alberta-Thy-11-98
\end{flushright}
\vspace{7mm}

\baselineskip= 6mm
\centerline{{\bf Yu. V. Gusev}$^{*\dagger}$ \
{\bf   and A. I.  Zelnikov}$^{*\ddagger}$}
\vspace{3mm}
\centerline{$^*${\em Theoretical Physics Institute, 
University of Alberta}}
\centerline{\em Edmonton, Alberta, Canada T6G 2J1}
\centerline{\em E-mails: ygusev, zelnikov@phys.ualberta.ca}
\vspace{3mm}   
\centerline{$^\dagger${\em Canadian Institute for Theoretical Astrophysics}} 
\vspace{3mm}   
\centerline{$^\ddagger${\em P.N. Lebedev Physics Institute}}
\centerline{\em Leninskii prospect 53, Moscow 117 924 Russia}
\vspace{10mm}

\centerline{\bf Abstract}  
           
\noindent
{\small	
Massless and massive scalar fields
and massless spinor fields are considered  at arbitrary 
temperatures  in four dimensional ultrastatic curved spacetime.
Scalar models under consideration can be 
either conformal or nonconformal
and include selfinteraction.
The one-loop nonlocal effective action at finite temperature  
and  free energy for  these quantum fields
are found up to the second order in background field strengths  
using the covariant perturbation theory.     
The resulting expressions are free of infrared divergences.  
Spectral representations for nonlocal terms
of high temperature expansions are obtained.}\\
\vspace{5mm}


\thispagestyle{empty}
\pagebreak
\baselineskip= 8mm
		
			\section{Introduction}                                             


Finite temperature field theory has been 
developed in a series of seminal papers
\cite{Matsubara59,Fradk59,MartSchwin-PR59}. 
Nowadays it is an actively growing branch of 
theoretical physics 
\cite{Kapusta-book89}.
Thermodynamical properties of thermal  quantum fields 
in the presence of background fields  are very important
for a large number of applications in high energy physics, 
astrophysics,  and cosmology.                                        
However most of these studies are devoted to
the situation when background fields  are  constant  (homogeneous)
\cite{DolJack-PRD74,MossTomsWright-PRD92}. This particular form 
of the effective action,  the effective potential 
\cite{Jackiw-PRD74,Avram-JMP95}, when large background fields
are taken into account nonperturbatively, 
is useful for study 
of  phase transitions in the early 
Universe or  quark-gluon plasma. 
For a long time, the opposite situation,
when  background fields are small but rapidly fluctuating, 
lacked  investigation  even in zero temperature  field theory. 
Traditional tools of quantum field theory, like the short proper  time
Schwinger-DeWitt expansion 
\cite{DeWitt-book65,BarVilk-PRep85,Ball-PRep89},
are intrinsically local, hence, they miss nonlocal contributions.
As a consequence of this deficiency artificial infrared divergences 
appear in the perturbative effective action for massless fields, and
perturbation theory breaks down. 
Finite temperature effects also contribute to  infrared 
divergences \cite{Kapusta-book89}, and
methods of diagram summations have been 
developed  to improve the perturbation series   
\cite{BraatPisar-PRD92,DrumHorLandReb-PLB97}.

To deal with massless field theories  properly,
 such as gauge field theories or quantum gravity,
Vilkovisky suggested  a new powerful method \cite{Vilk-Gospel} 
which is known as the covariant perturbation theory 
\cite{CPT1,CPT2,CPT3,CPT4}.
In these papers it was shown
that infrared divergences are artificial and brought into existence
by a mode of calculation 
rather than by a field  theory. 
They disappear after  summation of terms 
with infinite number of derivatives acting on background  fields, 
which results 
in  nonlocal terms entering the effective action \cite{CPT1}.
Such a summation can only be performed
in a given order in background field strengths.

Thermodynamics of  an ensemble of quantum fields 
in equilibrium in static curved  spacetimes 
is well defined, and most  properties of the system can 
be derived from its free energy 
\cite{DowkKenn-JPA78,Furs-hepth9709,FrolFurs-hepth9802}.  
In this paper we consider ensembles of  scalar 
and spinor fields in  the presence of external ultrastatic 
gravitational field.
The scalar models may have an arbitrary interaction potential
and an arbitrary coupling to gravity.    
We employ the method of covariant perturbation theory 
to find the finite temperature effective action 
and the corresponding free energy of these quantum fields 
on highly inhomogeneous gravitational backgrounds. 
An example of the situation when finite 
temperature effects on curved  background 
are  important, and, thus, nonlocal effective action is needed, 
is the Hawking radiation  by  black holes 
\cite{FrolVilk-PLB81,Vilk-CQG92}.

The paper is organized as follows. In the next  section we  
describe how to obtain  nonlocal free energy 
at finite temperature with help of the covariant perturbation theory.
In section~\ref{massless} we derive the free energy 
of interacting massless scalar fields
and study its high  temperature behavior.  
Massive scalar fields at high temperatures 
are treated in section~\ref{massive}.
In section~\ref{spinor} we derive  free energy 
for massless spinor  fields at finite and  high temperatures.                                       
Conclusion and discussion of possible applications 
and extensions of obtained results 
can be found in section~\ref{conclusion}.
We place necessary  complicated computations
into appendices~\ref{B} and \ref{C}.


\section{One-loop effective action and free energy of  
quantum  fields   in  ultrastatic   spacetimes}  

\label{ultrastatic}


Let us consider  fields 
$\varphi$ described by the classical action $S(\varphi)$ and 
the corresponding canonical
Hamiltonian in a generic curved static spacetime. 
Statistical free energy $F_{S}$ of the ensemble is defined 
as the trace of  logarithm of eigenvalues of 
the normal-ordered Hamiltonian.
In canonical quantization scheme, 
ultraviolet divergencies are traditionally subtracted 
from $F_S$ by the  normal ordering prescription 
\cite{FrolFurs-hepth9802,Allen-PRD86}. 
On the other hand,
in the imaginary time formalism of Matsubara \cite{Matsubara59}  
the problem of finding  free energy 
of the system in equilibrium  reduces to the computation of
the path integral of the Euclidean quantum field theory
and the corresponding effective action $W^{\beta}[g]$
\begin{equation}
	{\mathrm e}^{-W^{\beta}[g]}=
    \int D[\varphi]\, {\mathrm{e}}^{-S  [ \varphi, g] }.
\end{equation}
The temperature $T=1/(k_{B} \beta)$ enters the calculation via 
the condition of (anti)periodicity in the
Euclidean time $\tau$ imposed on quantum fields 
with (fermi) bose statistics
	\begin{equation}
	\varphi(x,\tau)=\pm\varphi(x,\tau+\beta)     \label{fields}
	\end{equation}
(the Botzmann's constant $k_B=1$  everywhere).

The canonical free energy $F_S$ and 
the thermal renormalized Euclidean effective action 
$W^{\beta}$ are closely related to each other and
differ \cite{DowkKenn-JPA78,Allen-PRD86} 
only by terms $\tilde{F}$ that are independent of temperature 
	\begin{equation}
	\frac{1}{\beta}W^{\beta} [\phi] 
    = F_S^{\beta}[\phi] + \tilde{F}[\phi],   
	\end{equation}
where $\phi=<\varphi>$ are mean fields.
The effective action is usually 
regularized using covariant methods, 
e.g., zeta-function \cite{DowkCrit-PRD76,Elizal-book95}, 
dimensional \cite{BarVilk-PRep85}, etc., while the canonical
free energy is regularized via normal
ordering of operators. The difference $\tilde{F}[\phi]$ is related 
to different ways of taking
into account vacuum energy contributions 
in covariant and canonical regularization schemes.   
The covariant approach is more appropriate
to our problems since it is consistent with calculations of 
the stress tensor and vacuum polarization effects in external fields.  
In any case, it is easy to compute $\tilde{F}(\phi)$
which is temperature independent and local \cite{Allen-PRD86}. 
Henceforth, we restrict our consideration to calculation 
of the one-loop Euclidean  effective action $W^{\beta}$
and the corresponding covariant Euclidean free energy 
\begin{equation}
	F^{\beta} \equiv \frac{1}{\beta} W^{\beta}. 
\end{equation}

We calculate  the free energy of quantum fields   on static 
background fields which include mean field $\phi$ 
and static gravitational field.
The Tolman temperature of such a  field  system in equilibrium 
is not constant throughout the static space. 
It is more  convenient to perform calculations of temperature 
effects in the Euclidean ultrastatic (optical)  spacetimes,
\begin{equation}
	{\mathrm d} s^2=
	{g}_{\mu\nu} {\mathrm d} x^\mu {\mathrm d} x^\nu
	= {\mathrm d} \tau^2+ 
{g}_{ij} {\mathrm d} x^i  {\mathrm d} x^j~, \label{ultra}
\end{equation} 
where local temperature is constant throughout the space. 
Ultrastatic and static spacetimes are related to each other by
a conformal transformation  of the metric.
Conformal properties of  the effective action
have  been studied in detail 
\cite{conform-trans},
and applied to free energy calculations by
Dowker and Schofield
\cite{DowkSch-PRD88,DowkSch-NPB89}. 
Using scaling properties of the finite temperature zeta 
functions it was shown  that  the difference of free energies 
in two conformally related
spaces does not depend on temperature.  
Then, all  temperature dependent terms can be
found from the free energy  in an ultrastatic  space,
where solution of the problem simplifies significantly.
This difference can be found by integrating the conformal anomaly, but
the method of Ref.~\cite{DowkSch-PRD88}
works for generic nonconformal  operators as well. 
 
The brief outline  of our research program  is 
to calculate free energy in an ultrastatic  space, and then
using a relation between  free energies in static and ultrastatic  spaces 
to express the final result  in terms of quantities 
defined in a physical (static) spacetime. 
In this paper we implement the first and most complicated  step
of obtaining $W^{\beta}$ and $F^{\beta}$ 
on the ultrastatic metric (\ref{ultra}).

Let us  consider  quantum $n$-component scalar field 
$\varphi\equiv\varphi_A$, \ $i =1 \dots n$, which satisfies 
the equation,
	\begin{eqnarray}
	[(\Box-{1\over 6}R)\hat{1} 
    + \hat{P}(\phi)] \varphi=0,  \label{1}
	\end{eqnarray} 
Our notations correspond to those of Refs.~\cite{BarVilk-PRep85,CPT2}:
the Laplacian $\Box= g^{\mu\nu} \nabla_{\mu} \nabla_{\nu}$
is constructed of covariant derivatives which are characterized
by the commutator curvature
\begin{equation}
(\nabla_{\mu} \nabla_{\nu}-
\nabla_{\nu} \nabla_{\mu} ) \varphi =
\hat{\cal R}_{\mu\nu}    \varphi.
\end{equation}            
This quantity is, of course, zero for scalar fields,
but we will need it in section~\ref{spinor} 
where spinor fields are considered.
The potential $\hat{P}$ may depend on the metric and mean field
$\phi$, which is a part of classical background. 
Thus, this class of models includes self-interacting fields.  
The vanishing potential $\hat{P}=0$ corresponds to the case 
of free conformal scalar fields, and  $\hat{P}= \hat{1}R/6$ -
to the minimally coupled free scalar fields.     
The overhat symbol indicates the matrix structures,  $\hat{P}={P^A}_B$,
and term $ R/6$ in  (\ref{1}) is 
explicitly singled out for convenience.  
The three field strengths $R_{\mu\nu},\ \hat{\cal R}_{\mu\nu},\  \hat{P}$   
will be  also referred to as  curvatures.
This  massless field theory will
be generalized to the case of massive fields  
in section~\ref{massive}. 

The one-loop Euclidean effective actions $W^{\beta}$  
is defined in terms of   the functional trace of the heat kernel,
\begin{equation} 
	 {W}^{\beta}= 
	\mbox{}-\frac1{2}  
	\int_0^{\infty} \frac{\mathrm{d} s}{s}  
	{\mathrm{Tr}}  {K}^{\beta} (s), 
\end{equation} 
where the heat kernel $\hat{K}^{\beta}(s)$ 
is the periodic in Euclidean time solution of the problem
\begin{equation}
	\left\{
	{{\mathrm d}  \over{\mathrm{d}} s} -
	\Big[\hat{1}(\Box-{1\over 6}R(x))+\hat{P}(x)\Big]
	\right\}
	\hat{K}^\beta (s| x,x')
	=\hat{1} \delta (s)  \delta(x,x'),         \label{heateq}
\end{equation} 
\begin{equation}
    \hat{K}^\beta(s|\tau,\mbox{\boldmath $x$};\tau',\mbox{\boldmath $x$}')
    =\hat{K}^\beta(s|\tau+\beta,\mbox{\boldmath $x$};\tau',
    \mbox{\boldmath $x$}').
\end{equation}      
The functional trace is understood as
\begin{equation}
{\mathrm{Tr}} K (s)  =
\int {\mathrm d}^D  x \,   
	{\mathrm tr}\,  \hat{K} (s), 
\end{equation}                                   
with the ${\mathrm tr}$ standing for the matrix trace, e.g., 
${\mathrm tr}\hat{1}= \delta^A{}_A$,
${\mathrm tr}\hat{P}=P^A{}_A$.

The thermal (periodic in the Euclidean time)   
heat kernel ${K}^{\beta}$ can be expressed  
as an  infinite sum of zero temperature (vacuum) heat kernels
\cite{DolJack-PRD74,DowkCrit-PRD77}
\begin{equation}
	\hat{K}^{\beta}(s|\tau, \mbox{\boldmath $x$}; \tau', 
	\mbox{\boldmath $x$}')
	=\sum_{n=-\infty}^{\infty} 
	\hat{K}(s|\tau, \mbox{\boldmath $x$}; {\tau}'+\beta n, 
	\mbox{\boldmath $x$}')         \label{sumK}
\end{equation}
This  image sum  is equivalent to summation
over Matsubara frequencies in a momentum space representation
in thermal field theory.
The image sum in the context of Casimir energy calculations
was introduced in Ref.~\cite{BrownMacl-PR69}.                    

Temperature effects are inherently  
connected  to the imaginary time.
It is convenient to factorize the heat kernel  
into temporal and spatial $K^{(3)}$ parts, 
\begin{equation} 
	 \hat{K}(s|\tau, \mbox{\boldmath $x$}; \tau',  
	\mbox{\boldmath $x$}') 
	= \frac{1}{(4\pi s)^{1/2}} 
    \exp \Big(  {-\frac{(\tau - {\tau}')^2}{4s}}  \Big) 
	 \hat{K}^{(3)}(s|\mbox{\boldmath $x$};   
	\mbox{\boldmath $x$}'),         \label{splitK} 
\end{equation} 
which is possible to do in ultrastatic spacetimes.       
Then, the trace of the heat kernel  takes a  form  
\cite{DowkKenn-JPA78},  
	\begin{eqnarray} 
	{\mathrm  Tr} {K}^{\beta}(s)&=&  
	\theta_3  
	\Big(0, 
 	 {\mathrm{e}  }^{-\frac{\beta^2}{4s}}  
	\Big)  
	\frac{\beta}{(4\pi s)^{1/2}}  
	\int {\mathrm  d}^3 x \,   
	{\mathrm{tr}}\,  \hat{K}^{(3)} 
	(s| \mbox{\boldmath $x$}, \mbox{\boldmath $x$}),
	\label{thetaTrK} 
	\end{eqnarray} 
when expressed in terms of the Jacobi  theta function \cite{GradRyzhik}, 
which is  defined in a usual way, 
	\begin{equation} 
	\theta_3 (a,b)  \equiv   \sum_{n=-\infty}^{n=\infty} 
	{{\mathrm{e}}  }^{2 n a{\mathrm{i}}} b^{n^2}.  \label{theta3} 
 	\end{equation}

The free energy of quantum  fields in static spacetime 
$ F^{\beta}$ is defined via   
the finite temperature Euclidean effective action
$ {W}^{\beta}$ and can be written in the form, 
\begin{equation} 
	 {F}^{\beta} = 
	\mbox{}-\frac1{2\beta}  
	\int_0^{\infty} \frac{\mathrm{d} s}{s}  
	{\mathrm{Tr}}  {K}^{\beta} (s). 
\end{equation} 
 
The vacuum mode $n=0$ in  the infinite sum 
(\ref{thetaTrK})-(\ref{theta3})
corresponds to the zero temperature  effective action which 
suffers ultraviolet divergencies 
\cite{DeWitt-book65,BarVilk-PRep85,Elizal-book95}. 
Fortunately, this is the only divergent term of the sum 
\cite{DowkKenn-JPA78}, 
so it is convenient to treat it separately. 
We subtract  the zero temperature ($\beta=\infty$) free energy 
$ {F}^{\infty}$ from $ {F}^{\beta}$ 
and  renormalize it with the  use of  the zeta function regularization 
\cite{DowkCrit-PRD76,Elizal-book95,Gibbons-PLB77}, 
\begin{equation} 
	 {W}^{\infty}_{\mathrm{ren}}= 
	\mbox{}-\frac12 \frac{\partial}{\partial\epsilon} 
	\left[ 
	\frac{\mu^{2\epsilon}}{\Gamma(\epsilon)} 
	\int_0^{\infty}\! \frac{{\mathrm{d} }s}{s^{1-\epsilon}}\,  
	{\mathrm{Tr}}  {K}(s)  
	\right]_{\epsilon=0},         \label{zeta-reg} 
\end{equation} 
where $\mu$ is a mass-like regularization parameter 
and $\Gamma$ is the gamma function.                           
$ {F}_{\mathrm{ren}}^{\infty}$  will  be combined with $n \neq 0$  
terms at the end of our derivations. 
Therefore, we compute 	
\begin{equation} 
	 {F}^{\beta}_{\mathrm{ren}} -  {F}^{\infty}_{\mathrm{ren}} = 
	\mbox{}-\frac{1}{2}    
	\int_0^{\infty}\! \frac{{\mathrm{d} } s }{s} 
	\Big(    \theta_3  
	\big(0,{\mathrm{e}}^{-\frac{\beta^2}{4s}}\big) -1 \Big)        
	\frac{1}{(4 \pi s)^{1/2}} 
	\int {\mathrm{d}}^3 x \,  
	{\mathrm{tr}}  \hat{K}^{(3)}  
( s | \mbox{\boldmath $x$}; \mbox{\boldmath $x$}  ).     \label{Wb-W} 
\end{equation}                 
The  heat kernel $ {K}^{(3)}(s)$ is defined as a solution 
of Eq.~(\ref{heateq})  with the three dimensional operator, 
\begin{equation} 
	\left\{
	{{\mathrm d} \over{\mathrm{d}}s}
	-\Big[\hat{1} (\triangle -{1\over 6}R(\mbox{\boldmath $x$}) )  
	+  \hat{P}(\mbox{\boldmath $x$})\Big]
	\right\}
	\hat{K}^{(3)}(s|\mbox{\boldmath $x$};\mbox{\boldmath $x$})=
\hat{1}\delta(s)\delta(\mbox{\boldmath $x$},\mbox{\boldmath $x$}').  
	\label{oper3d} 
\end{equation}  
In this case the three dimensional Laplacian $\triangle$,  
and  potential  $ \hat{P}(\mbox{\boldmath $x$})$ and  
 the curvature $ R_{ij}(\mbox{\boldmath $x$})$  
are defined on a three dimensional hypersurface \
$\tau={\mathrm{const}}$ of the ultrastatic spacetime.  
 
Many methods  have been developed  for calculation of the trace of  
the heat kernel 
\cite{DeWitt-book65,SDWcoeff}.  
Most of them (see reviews \cite{Ball-PRep89,BarVilk-PRep85}) 
reduce to various representations of its  small $s$ expansion,
\begin{equation}
    {\mathrm  Tr} K(s)=\frac1{(4\pi s)^{D/2}} 
    \int {\mathrm d} x^D g^{1/2}(x) \,
\sum_{n=0}^{\infty} s^n  {\mathrm tr} \,\hat{a}_n (x,x), \ \ \
s \rightarrow 0. \label{SDW}
\end{equation} 
However, as soon as the inverse temperature
$\beta $ is finite, the behavior of the heat kernel at large values
of proper  time $s$ becomes very important  \cite{DowkKenn-JPA78}.
Therefore, expansion (\ref{SDW})  is not suitable for our task 
of finding the free energy at finite temperature.
Besides, Schwinger-DeWitt coefficients $a_n$ are local functions
of background fields, henceforth, 
nonlocal free energy cannot be derived using (\ref{SDW}).
To solve the problem  of obtaining  
nonlocal free energy at finite temperature
we have to resort  to  the covariant perturbation theory  
\cite{CPT1,CPT2,CPT3,CPT4}. 
There is no  need to repeat derivations 
of the covariant perturbation 
theory here because an expression for ${\mathrm Tr}\, K (s)$  
is already known in arbitrary $D$-dimensions \cite{CPT2,CPT4}.  
In this paper we will take it  up to terms quadratic in curvatures, 
\begin{eqnarray} 
{\mathrm{Tr}} K(s) &=& \frac1{(4\pi s)^{D/2}}\int\! 
    {\mathrm{d}}^{D}x\, 
	g^{1/2}\, {\mathrm{tr}}\, \Big\{  \hat{1} +  s \hat{P} 
\nonumber\\[2mm]&& \mbox{} 
	+s^2 \Big[ 
	R_{\mu\nu} f_{1}(-s\Box) R^{\mu\nu}   \hat{1} 
	+ R f_{2}(-s\Box) R \hat{1}   
    + \hat{P} f_{3}(-s\Box) R  
\nonumber\\[2mm]&& \mbox{} 
		+ \hat{P}  f_{4}(-s\Box) \hat{P} 
    + \hat{\cal R}_{\mu\nu} f_{5}(-s\Box) \hat{\cal R}^{\mu\nu}
	 \Big]\Big\} 
	+{\mathrm{O}}[\Re^3].  \label{TrK} 
\end{eqnarray} 
Analytic functions $f_i $ ({\em form factors})  
have the dimensionless argument  $s\Box$.
(The appearance of nonlocal form factors in the momentum space
representation of the effective action
originates in the classical paper of  Schwinger  \cite{Schwin-PR51}).
The  form factors  act on tensor invariants constructed of 
the set of field strengths   
$R^{\alpha\beta},\  \hat{P}, \  \hat{\cal R}_{\mu\nu}$ 
characterizing  background.  
The collective notation $\Re$ will be  used  for these curvatures.  
First two terms of the sum (\ref{TrK}) are purely local and coincide 
with first two coefficients of the short  proper  
time expansion (\ref{SDW}). 
Formally, the expansion (\ref{TrK})  is valid only in  
an asymptotically flat  Euclidean spacetime 
with the topology $R^{D}$. 
All background curvatures  $\Re$ are supposed to
vanish at spacetime infinity \cite{CPT2}. 
Since we use a  perturbation theory, 
all calculations  are carried out with accuracy 
${\mathrm{O}}[\Re^n]$, i.e., 
up to terms of $n$-th and higher power in the curvatures $\Re$. 
The very structure of this curvature expansion  
restricts its validity to background fields satisfying the relation,  
\begin{equation} 
	\nabla \nabla \Re >> \Re^2. 
\end{equation} 
Physically it means that gravitational  fields are small
in magnitude but quickly oscillate.

All form factors in (\ref{TrK}) can be   
expressed in terms of one  basic form factor 
\begin{equation} 
	f(-s\Box)=\int_0^1\!\! {\mathrm{d}}\alpha\:  
	{\mathrm{e}}^{\alpha(1-\alpha)s\Box}. \label{basicf} 
\end{equation}  
Their explicit form reads \cite{CPT2} 
\begin{eqnarray} 
	f_1  (-s\Box) &=& 
    \frac{f(-s\Box)  -  1-  \frac16   s\Box}{(s\Box)^2},    
\label{hff1}\\ 
	f_2  (-s\Box) &=& \frac18   \left[ 
  	\frac1{36}\,  f( -s \Box )  -  
    \frac13\,  \frac{  f( -s \Box ) - 1}{ s \Box }- 
  	\frac{  f( -s \Box ) - 1 - \frac16 s \Box }{ (s \Box)^2} 
				\right], 
\label{hff2}\\ 
	f_3( -s \Box ) &=& 
  \frac1{12}\,   f( -s \Box ) - \frac12\,   \frac{f( -s \Box )-1}{ s \Box },  
\label{hff3} \\     	
	f_4   ( -s \Box ) &=& \frac12  \,  f( -s \Box ), 
 \label{hff4} \\
	f_5   ( -s \Box ) &=& \frac12  \, \frac{f( -s \Box )-1}{ s \Box }. 
 \label{hff5}
\end{eqnarray}          
Even though, in the following consideration  general covariance
is broken because of the presence of  temperature, 
we will refer to this curvature expansion as to the  covariant 
perturbation theory.  
In spatial dimensions the covariance remains explicit.

A few words about validity of this approximation are in order. 
Since we consider quantum fields at some fixed temperature,  
one can  say that  the field  system in question  represents 
a canonical ensemble.
To define a canonical ensemble rigorously we have to assume that
the fields  are in some cavity of a finite volume, as it is usually assumed
in  the presence of a black hole \cite{FrolFurs-hepth9802}.
This assumption should be reconciled, however, with our method
of computation described above, which in the present form works
only for asymptotically flat spacetimes and  requires
vanishing  background fields at spacetime infinity.
It is  important  to  note that background field strengths, 
sources of vacuum polarization,
have a compact support on a manifold, thus, providing
an effective volume cut-off.   In regard to gravitational field
this property is due to the presence of the Ricci tensor
rather than  the Riemann tensor  \cite{CPT2}.


               \section{Free energy of massless scalar fields}

\label{massless}      


Let us now compute free energy (\ref{Wb-W}) of massless scalar 
fields at  finite temperature.
This case was briefly reported  
in our letter \cite{GusZel-CQG98}.
After introducing a new variable $y =\beta^2/4 s$,
first two terms  of the trace of the heat kernel (\ref{TrK})
take the form of the integral,
\begin{equation}
	\int_{0}^{\infty}{\mathrm{d}} y   \,
	\left[\theta_3\Big(0,{\mathrm{e}  }^{-y}\Big)-1 \right] \,  y^{a-1}
	= 2 \, \zeta (2 a) \,  \Gamma(a),          \label{local}
\end{equation}
where $\zeta$ is the Riemann zeta function,
$\Gamma$ is the  gamma function.
When  $a$ is taking values 2 and 1, expression (\ref{local})
gives for the zeroth and first curvature orders
coefficients $\pi^4/45$ and $\pi^2/3$ correspondingly.
These local contributions  to the free energy are well known
 \cite{DowkKenn-JPA78,BarFrolZel-PRD95}
and coincide with the first two terms of high temperature expansion.
 Since all information about temperature is 
separated from tensor invariants, we  can write 
down an anticipated form of free energy up to second order 
in the field strengths,                                                                 
\begin{eqnarray}
	 {F}^{\beta}_{\mathrm{ren}} -  {F}^{\infty}_{\mathrm{ren}}&=&
	\mbox{}- 
	\int \!  {\mathrm{d} }^3 x \,  {g}^{1/2}\, 
        	{\mathrm tr} \Big\{
	\frac{\pi^2}{90 \beta^4} \hat{1}+
	 \frac{1}{24 \beta^2} \hat{P}  
\nonumber\\ && \mbox{}
	+  \frac1{32\pi^2} \Big[\,
	R_{i j} \gamma_1^{\beta}(-\triangle) R^{i j} 
	+ R \gamma_2^{\beta}(-\triangle) R 
\nonumber\\ && \mbox{}
	+ \hat{P} \gamma_3^{\beta}(-\triangle) R 
	+ \hat{P} \gamma_4^{\beta}(-\triangle) \hat{P}   \, \Big]
	        +{\mathrm{O}}[{\Re}^3]\Big\}.  \label{efacT}
\end{eqnarray}

Then,  the problem with the second curvature order
is reduced now to calculation of  the thermal form factors, 
\begin{equation}
	\gamma_{i}^{\beta}({-\triangle}) \equiv 
    \gamma_{i} ({\beta} \sqrt{-\triangle}) =
	\int_0^\infty {{\mathrm{d} } s \over s}
	\left[\theta_3  
\Big( 0,  {\mathrm{e}  }^{-\frac{\beta^2}{4 s}}\Big) - 1 \right] 
	f_{i}   (  - s  \triangle ), \label{gammaTi}
\end{equation}   
where $f_i$ are given by (\ref{hff1})--(\ref{hff4}).
We show  how to compute  (\ref{gammaTi}) 
when $f_i (-s \triangle)$      is the basic form factor  (\ref{basicf}). 
After substituting (\ref{basicf})
 into (\ref{gammaTi})  and writing down 
the theta function  (\ref{theta3})  explicitly we get,
\begin{equation}
	\gamma (\beta  {\sqrt{ -\triangle}}) =
     2 \sum_{n=1}^{\infty}
	\int_0^1 {\mathrm{d} }\alpha 
	\int_0^{\infty} \frac{ {\mathrm d} y}{y}
\exp  \Big( - y n^2 -  \frac{1}{4 y} \alpha (1-\alpha) 
\beta^2 (- \triangle ) \Big).
\end{equation} 
Integration  over $y$ produces  
the  modified Bessel function of the second kind
\begin{equation}
	\gamma (z)= 
	4 \sum_{n=1}^\infty 
	\int_0^1 {\mathrm{d} } \alpha
	K_0  \left( n z \sqrt{\alpha(1-\alpha)}\right),  \label{Bessel0}
\end{equation}                                                    
where $  z= \beta  {\sqrt{ -\triangle}} $.  
Change of variables, $x = 2 \sqrt{\alpha(1-\alpha)}  $,  
allows us to express (\ref{Bessel0})  in terms of 
the exponential integrals, 
\begin{equation}
	\int_0^1 {\mathrm d}x {  x  \over  \sqrt{1-x^2}  }
    K_0   (  \frac{nz~x}{2}  )
	={1\over n z}\Big[\,   { \mathrm Ei}\big(\frac{nz}{2}\big) \,  
	 {\mathrm e}^{-{nz}/{2}}
	-   {\mathrm Ei}   \big(-\frac{nz}{2}\big) \,
	 {\mathrm e}^{{nz}/{2}}\, \Big].        \label{Ei}
\end{equation}  
Now we can use for the right hand side of (\ref{Ei}) 
its standard form in terms of elementary functions \cite{GradRyzhik} 
and obtain
\begin{equation}
	\gamma (z)=
	4 \int_0^\infty \! {\mathrm{d} }t \, \sum_{n=1}^\infty 
	\,{\sin(t)\over  t^2 + n^2  z^2 /4}.  \label{sinus}
\end{equation}  
The sum over $n$ can be evaluated \cite{GradRyzhik}, 
\begin{equation}
\sum_{n=1}^{\infty} \frac{1}{t^2+n^2 z^2 /4}=
\frac12 \left[\frac{2\pi}{zt}\,\frac{1}{{\rm th}(2\pi t/z)}
-\frac{1}{t^2}\right],    \label{mainsum}
\end{equation}
and the resulting expression reads, 
\begin{equation}
	\gamma ( z ) = 
	2 \int_0^\infty \!  {\mathrm{d}} t \,
	\sin(t)
	\left[
	\frac{2\pi}{zt}\,    \frac{1}{{\rm th}(2\pi t/z)}   
	- \frac{1}{t^2}     \label{hgamma0}
	\right].
\end{equation}       

As can be seen from (\ref{hff1})--(\ref{hff4}), there are two other types of 
basic thermal form factors  (with one and with two subtractions).
Their derivations can be found in appendix A. 
Applying  results  (\ref{hgamma0}),  (\ref{hgamma1}), 
and (\ref{hgamma2}) to the table
of form factors  we obtain for  all  thermal 
form factors the following expression,
\begin{equation}       
	 \gamma_i ( \beta \sqrt{-\triangle}) = 
    \int_0^\infty\!\! {\mathrm{d} }t \,\, g_i(t)
	\left[ 
	\frac{2\pi}{ \beta  {\sqrt{ -\triangle}}\,  t}\, \frac{1}{{\mathrm th}
	(2\pi t/(\beta{\sqrt{ -\triangle}}))} 
	- \frac{1}{t^2}    \label{gammaTgen} \right],   
\end{equation}
and $g_i \ (i=1,...4)$\ are simple combinations of elementary functions
\begin{eqnarray}       
	 g_1(t)&=&  
	 \mbox{}-\frac12 \Big( \frac{\sin (t)}{t^2} 
	+ 3\Big[{\cos (t)\over t^3}-{\sin(t)\over t^4}\Big]\Big),
\\  
	 g_2(t)&=& \frac1{48}\Big(
	  \frac1{3} \sin (t)  + 2 {\cos (t)\over t} + {\sin(t)\over t^2}
	+ 9\Big[ {\cos (t)\over t^3}
	- \frac{\sin(t)}{t^4}\Big]\Big),
\\    
	  g_3(t)&=& \frac12\Big(
	  \frac1{3} \sin (t) +{\cos (t)\over t}
	- {\sin(t)\over t^2}\Big),  \label{gammasT}  
\\            
	g_4(t)&=& \sin(t).
\end{eqnarray}

The final result for renormalized free energy at finite temperature   
$ {F}_{\mathrm{ren}}^{\beta}$
is presented by a sum of   Eqs.~(\ref{efacT})--(\ref{gammasT}) 
and renormalized
free energy at zero temperature $ {F}^{\infty}_{\rm ren}$.
After the  zeta regularization (\ref{zeta-reg}), 
the latter one takes the  form,
\begin{eqnarray}  
         {F}^{\infty}_{\mathrm{ren}}& = &\mbox{}
	-  \frac1{32\pi^2}  \int\! 
	{\mathrm d}^3 x \,  {g}^{1/2}   
        {\mathrm tr}\Big\{ R_{i j } \gamma_1(-\triangle) R^{i j }
	+ R \gamma_2(-\triangle) R 
\nonumber\\ && \mbox{}
	+ \hat{P} \gamma_3(-\triangle) R
	+ \hat{P} \gamma_4(-\triangle) \hat{P}   
             +{\mathrm{O}}[\Re^3]\Big\}.       \label{efacz}
\end{eqnarray}
where zero temperature form factors 
$\gamma_i(-\triangle)$,\ $i=1, ... 4$, are
\begin{eqnarray}	
	&&\gamma_1(-\triangle)=
	\frac1{60} \left[-{\rm ln}\Big(-\frac{\triangle}{\mu^2}\Big)
	+\frac{46}{15}\right], 
\label{gamma1} 
\\ &&
	\gamma_2(-\triangle)=\frac1{180}   \left[
	{\rm ln}\Big(-\frac{\triangle}{\mu^2}\Big)-\frac{97}{30}\right],   
\label{gamma2} 
\\ && 
	\gamma_3(-\triangle)=\mbox{} - \frac1{18},   
\label{gamma3}
\\ &&
	\gamma_4(-\triangle)=
	\frac12 \left[-{\rm ln}\Big(-\frac{\triangle}{\mu^2}\Big) +  2 \right]. 
\label{gamma4}
\end{eqnarray}  
This expression differs from the one obtained 
using dimensional regularization
only by unessential additive constants \cite{CPT2}.

Formulae ({\ref{efacT})--(\ref{gamma4}), we have obtained,
are valid at arbitrary finite temperature. 
Now we would like to study asymptotic behavior of the free energy
(\ref{efacz}) in high temperature regime,
the most interesting and the best studied limit.
In the framework of perturbation theory, 
the problem boils down to finding 
 $\beta \rightarrow 0$ asymptotic of 
thermal form factors (\ref{gammaTgen}). 
We have to be careful while dealing with mutually compensating
singularities.
After relatively straightforward calculations  
the outcome for (\ref{hgamma0}) is, 
\begin{eqnarray}
{\gamma} ( \beta \sqrt{-\triangle}) &=&     
\frac{2\pi^2}{\beta \sqrt{-\triangle} }
+2\Big[
	{\rm ln}(\frac{\beta\sqrt{-\triangle}}{4 \pi})-1 +{\mathrm{C}}
\Big]
\nonumber\\&&\mbox{}
-\frac{\zeta(3)}{24\pi^2}\beta^2 (-\triangle) 
+\frac{\zeta(5)}{640\pi^4}\beta^4 (-\triangle)^2 + 
{\mathrm{O}}[\beta^6],
\ \ \ \beta \rightarrow 0,                             \label{basicgammaT}
\end{eqnarray}                  
where ${\mathrm{C}}$ is Euler's constant 
and $\zeta$ is the Riemann zeta function.

Now, expressions for the vacuum free energy (\ref{efacz})
and the high temperature expansion of   (\ref{efacT}) match,
and can be combined into a single formula. 
The resulting $T \rightarrow \infty$ expansion
of the renormalized one loop free energy takes a form,
\begin{eqnarray}
	 {F}_{\mathrm{ren}}^{\beta}&=&
	\mbox{}-\int\!{{\mathrm d}}^3 x\,  {g}^{1/2}\, 
        	{\mathrm tr}\left\{\frac{\pi^2}{90\beta^4}\hat{1}
	+\frac{1}{24\beta^2} \hat{P} \right.
\nonumber\\ && \mbox{} 
	+ \frac{1}{32\beta}  \Big[
	\frac{1}{16} R_{i j}\frac{1}{\sqrt{-\triangle}} R^{i j }
	-\frac{25}{1152} R\frac{1}{\sqrt{-\triangle}}R 
	-  \frac{1}{12}\hat{P}\frac{1}{\sqrt{-\triangle}}R
	+ \hat{P}\frac{1}{\sqrt{-\triangle}}\hat{P}\Big]  
\nonumber\\   && \mbox{}
	+\frac{1}{16\pi^2} \Big(  
	\ln (\frac{\beta\mu}{4\pi})+{\mathrm{C}}\Big) \,
	\Big[
	\frac{1}{60}  R_{ij} R^{ij}
	-\frac{1}{180} R  R 
	+ \frac12 \hat{P} \hat{P}\Big]  
\nonumber\\&&\mbox{}
	+\beta^2 \frac{\zeta (3)}{128 \pi^4} 
	\Big[
	\frac{1}{840}  R_{ij}\triangle  R^{ij}
	- \frac{1}{3780} R \triangle  R 
       + \frac{1}{180} \hat{P} \triangle R
	+ \frac1{12} \hat{P}  \triangle    \hat{P}\Big]                                 
\nonumber\\&&\mbox{}
	+\beta^4 \frac{3 \zeta (5)}{1024 \pi^6} 
	\Big[
	\frac{1}{15120}  R_{ij} {\triangle}^2   R^{ij}
	+ \frac{1}{1260} \hat{P}  {\triangle}^2  R 
	+ \frac1{120} \hat{P} {\triangle}^2  \hat{P}\Big] 
\nonumber\\&&\left. \mbox{}                 
	+{\mathrm{O}}[{\Re}^3]
	+{\mathrm{O}}[\beta^6] \vphantom{I}  \right\},
 	\  \  \beta \rightarrow 0. \label{efachighT}
        \end{eqnarray}                      
All local terms of this result  perfectly 
reproduce those of   Refs.~\cite{DowkKenn-JPA78,DowkSch-PRD88}.  
The combination of quadratic in curvatures terms at the logarithm
is just  the trace of the second Schwinger-DeWitt coefficient $a_2$, 
taken with Riemann curvature expressed via Ricci one 
\cite{DeWitt-book65,BarVilk-PRep85,CPT2}. 
Higher powers of $\beta$ in Eq.~(\ref{efachighT})
are also quadratic in  curvatures parts of 
$a_3$ and $a_4$  Schwinger-DeWitt coefficients
\cite{SDWcoeff,BGVZ-JMP94-asymp}.   

We obtained the explicit form of all nonlocal terms
of the second order in curvatures. 
They are  proportional to $1/\beta$, and were known to exist
 \cite{DowkSch-PRD88}.
The general structure  of nonlocal terms is 
$\Re \frac{1}{\sqrt{-\triangle}}\Re$,
and, therefore, techniques based on local (small $s$) expansions 
could not generate them.  Terms of higher orders in curvatures 
\cite{CPT4,BGVZ-JMP94-asymp}  will also
give nonlocal contribution linear in temperature.

The meaning  of nonlocal  structures
can be understood from spectral representations
in terms of massive Green functions \cite{Schwin-book89,CPT3,CPT4}.
For this particular form we have the following spectral formula,
\begin{equation}
\frac{1}{\sqrt{-\triangle}}=
\frac{2}{\pi}\int_{0}^{\infty}\! {{\mathrm d} }m \, 
\frac{1}{m^2-\triangle}.  \label{spectral}
\end{equation}                                                                                            
A remarkable  property of the expression (\ref{efachighT}) 
is that it contains the only kind of nonlocality, (\ref{spectral}). 
All logarithmic nonlocalities $\ln (-\triangle)$,
that are present in $F^{\infty}_{\mathrm ren}$  
and $F^{\beta}_{\mathrm ren}$,
have mutually canceled, leaving logarithm  
temperature dependance 
in the form of $\ln (\beta \mu)$,
This local combination is well known in both  flat   
\cite{DolJack-PRD74,HabWeld-PRD82}
and curved \cite{DowkKenn-JPA78} space thermal field theory. 
The $\ln (-\triangle)$ disappearance is still being analyzed
in a different physical language and in a different setting 
\cite{BranFrenk}.

Of course, we are not completely satisfied with the integral representation
for the free energy at finite temperature (\ref{efacT}). Although, it admits
a closed form, we would prefer  to see $F^\beta$ expressed entirely in terms
of analytical and special  functions. Indeed, it is possible to obtain
such a form after applying the Poisson resummation \cite{GradRyzhik},
\begin{equation}
    \sum_{n=-\infty}^{\infty} 
    {\mathrm{e}}^{- \frac{\beta^2}{4 s}n^2}  = 
    \frac{\sqrt{4\pi s}}{\beta}
    \sum_{ k = -\infty}^{\infty}   
    {\mathrm{e}}^{- \frac{4 \pi^2 s }{ \beta^2 } k^2 }.      \label{Poisson}
\end{equation}    
Then, the following identity holds,
	\begin{equation}
    \theta_3  \Big( 0,  {\mathrm{e}  }^{-\frac{\beta^2}{4 s}n^2}\Big) - 1 
    = \frac{\sqrt{4 \pi  s} }{ \beta }
	\left(   
	\sum_{k=-\infty}^{\infty}
	{\mathrm e}^{- \frac{4 \pi^2 s}{\beta^2} k^2}  
	- \int_{-\infty}^{\infty} {\mathrm d} \kappa \, \,
	{\mathrm e}^{-\frac{4\pi^2 s}{\beta^2} \kappa^2}
	\right).                         \label{newTheta}
	\end{equation}  
We compute now the basic thermal form factor, (\ref{gammaTi}) 
with (\ref{basicf}), using this identity and separating $k=0$ term 
out of the sum,
\begin{equation}
              \gamma (\beta\sqrt{-\triangle}) =
	\frac{2 \pi^2}{\beta \sqrt{-\triangle}} +
	\frac{8 \pi}{\beta \sqrt{-\triangle}}
	\left(   
		\sum_{k=1}^{\infty} {\mathrm arctan} 
		\Big( \frac{\beta\sqrt{\triangle}}{4 \pi k} \Big)  
    - \int_{0}^{\infty} \! {\mathrm d} \kappa\, {\mathrm arctan} 
		\Big( \frac{\beta\sqrt{\triangle}}{4 \pi \kappa} \Big)
	\right).		
\end{equation}                           
The $k=0$ mode of the  sum  gives precisely 
the leading  infinite temperature 
contribution, while the rest can be calculated
by employing the following sum,  
\begin{equation}
	\sum_{k=1}^{\infty}
 	\left( {\mathrm arctan} 
		\Big( \frac{b}{ k} \Big) 
	- \frac{b}{k}
	\right)   =  
	\frac{\mathrm i}{2}\ln \left(
	\frac{\Gamma (1 + {\mathrm i} b ) }{\Gamma (1 - {\mathrm i} b)}
	\right)      
		- b {\mathrm C}.                       	
\end{equation}
Adding up the regularized zero temperature form factor,  
\begin{equation}
           \gamma (-\triangle)
	= - \ln \Big( \frac{-\triangle}{\mu^2} \Big) +2, \label{zeroTff}
\end{equation}
we obtain an expression which is valid at any temperature,
\begin{equation}
\gamma (\beta \sqrt{-\triangle})=
\frac{2 \pi^2}{\beta \sqrt{-\triangle}} +
\frac{4 \pi {\mathrm i}}{\beta \sqrt{-\triangle}}
\ln \left(
\frac{\Gamma (1 + {\mathrm i} 
\frac{\beta \sqrt{-\triangle}}{4 \pi}) }{\Gamma (1 - {\mathrm i}
\frac{\beta \sqrt{-\triangle}}{4 \pi})}
\right)                                                      
+ 2 \ln \Big(  \frac{\beta \mu}{4 \pi} \Big).         \label{newffT}
\end{equation}

Besides an obvious advantage of Eq.~(\ref{newffT}), namely, 
that it is the formula  in terms of usual elementary and special functions,
the leading infinite  temperature 
contributions  are present here explicitly. 
Taking $\beta \rightarrow \infty$  and $\beta \rightarrow 0$ limits,
one can readily find zero temperature (\ref{zeroTff}) 
and high temperature 
(\ref{basicgammaT}) asymptotics of this basic thermal form factor.  
In fact, one can see that the logarithm of the gamma functions' ratio
in the main result (\ref{newffT})
is a sum of all positive powers of $\beta$ in   
the high temperature limit (\ref{basicgammaT}).          
Hence, it  gives a partial (in the given curvature order) summation 
formula for the $\beta \rightarrow 0$ 
series \cite{DowkKenn-JPA78}.
Eventually, one has to transform (\ref{newffT})
into a  spectral form,  the procedure we 
can complete again  only at high temperatures,
(\ref{spectral}). This is the reason why 
we refrain from deriving the total free energy 
in this new representation.


               \section{Free energy of massive scalar fields}    

\label{massive}


The use of the curvature expansion is crucial for derivation
of the massless field free energy, because it allows one to avoid 
artificial infrared divergences. 
Two other advantages of perturbation theory,
namely,  that free energy can be found at arbitrary finite temperature
and important  nonlocal contributions can be obtained,
work for a thermodynamic system of  massive fields as well. 
Besides, this is the most studied field model, so let us investigate 
an  ensemble of multi-component 
scalar massive fields satisfying equation,
\begin{equation}
     \Big[  \Big( \Box - \frac16 R  - m^2 \Big) \hat{1} 
+  \hat{P}(\phi) \Big] \varphi = 0.  
\end{equation}
Because the mass term can be factorized out of the heat kernel,
one can still use massless heat kernel (\ref{TrK}) to derive 
the free energy,
\begin{equation}
	{F}^{\beta}
		= - \frac{1}{2 \beta}
	\int_{0}^{\infty} \frac{{\mathrm d} s}{s}\,
	{\mathrm e}^{-s m^2}{\mathrm Tr} {K}^{\beta}(s).
\end{equation}
 
As usual, we subtract $n=0$ mode from the image sum, (\ref{Wb-W}).
Let us first treat local terms. The result in terms of 
the modified Bessel functions reads,
\begin{eqnarray}  
        {F}^{\beta}_{\mathrm{ren}}- {F}^{\infty}_{\mathrm ren}  &=&
	\mbox{}- \frac{1}{32 \pi^2}
	\int\!  {\mathrm d}^3  x\,  {g}^{1/2}   
	\sum_{n=1}^{\infty} {\mathrm tr}  
	\left\{ 		\Big( 	\frac{16 m^2}{ \beta^2 n^2} K_0(m \beta n)   
		+ \frac{16 m}{ \beta^3 n^3} K_1(m \beta n)
	\Big)  \hat{1} \right. 
\nonumber\\&&\left.\mbox{}
	+ \frac{8 m}{ \beta n} K_1(m \beta n)\hat{P} 
	+ {\mathrm O}[\Re^2]\right\}. 
\end{eqnarray}
So far this  expression is valid at any nonzero temperature.
However, we  are able to proceed 
and obtain explicit formulae  only in  high temperature limit.
Simple expansions of the Bessel functions at $\beta \rightarrow 0$
with the subsequent \hbox{$n$-sum} evaluation produces known local
contributions \cite{DowkSch-PRD88,DowkSch-NPB89}. 
The total result for free energy of massive fields 
at high temperature looks like:
\begin{eqnarray}
	{F}^{\beta}_{\mathrm{ren}} - {F}^{\infty}_{\mathrm{ren}}&=&
	\mbox{}- 
	\int \!  {\mathrm{d} }^3 x \, {g}^{1/2}\, 
        	{\mathrm tr}\left\{
	\frac{\pi^2}{90 \beta^4}\hat{1}+
	 \frac{1}{24 \beta^2} (\hat{P}  - m^2 \hat{1})
\right.
\nonumber\\ && \mbox{}	
               +   \frac{1}{32\pi^2}  \Big[
	R_{i j } \gamma_1^{\beta}(-\triangle) R^{i j } \hat{1}
	+ R \gamma_2^{\beta}(-\triangle) R\hat{1} 
\nonumber\\ && \left. \mbox{}
	+ \hat{P} \gamma_3^{\beta}(-\triangle) R
	+ \hat{P} \gamma_4^{\beta}(-\triangle) \hat{P} \Big]   
    +{\mathrm{O}}[\Re^3]  \right\}, \  \   
\beta \rightarrow 0,   \label{efactTmass}
\end{eqnarray}

The computational procedure for second order terms
is performed  after Poisson resummation (\ref{Poisson}).   
Applying Eq.~(\ref{newTheta}) to the basic form factor 
of nonlocal free energy for massive fields ,
\begin{equation}
    \gamma^{\beta}({-\triangle}) =
    \int_0^\infty {{\mathrm{d} } s \over s}
    \left[\theta_3  
\Big( 0,  {\mathrm{e}  }^{-\frac{\beta^2}{4 s}}\Big) - 1 \right] 
    {\mathrm e}^{-s m^2} f  (  - s  \triangle  ),       \label{basicffM}
\end{equation}  
(vacuum contribution  subtracted in (\ref{basicffM}) 
is dealt with at the end of the present section),
and using the integral
\begin{equation} 
    \int_0^1 {\mathrm d} \alpha 
    (m^2 -\alpha (1 - \alpha) \triangle)^{(-1/2)}
    = \frac{2}{\sqrt{-\triangle}} 
 {\mathrm arctan} 
\Big( \frac{\sqrt{-\triangle}}{2 m } \Big) \label{alpha-int}
\end{equation}
we get                                                  
\begin{eqnarray}
    \gamma (\beta\sqrt{-\triangle})& =&   
    \frac{ 4   \pi    }{  \beta   \sqrt{   - \triangle}    } \
    {\mathrm arctan}
	\left(
      	\frac{  \sqrt{-  \triangle} }{ 2 m  }     
	\right)
\nonumber\\ && \mbox{} 
	+\frac{8 \pi}{\beta \sqrt{-\triangle}}
	\left[   
		\sum_{k=1}^{\infty} {\mathrm arctan} 
	\left( \frac{\sqrt{-\triangle}}{
		\sqrt{4 m^2 + {16 \pi^2 k^2}/{\beta^2} }} \right) \right.
\nonumber\\ &&\left. \mbox{} 
		- \int_{0}^{\infty} 
	\! {\mathrm d} \kappa\, {\mathrm arctan} 
	\left( \frac{\sqrt{-\triangle}}{
		\sqrt{ 4 m^2 + {16 \pi^2 \kappa^2}/{\beta^2} }} \right)
	\right].	\label{gammaffm}	
\end{eqnarray} 
This equation  is valid at arbitrary finite temperature, therefore,
free energy of massive fields is  nonlocal at any temperature.
The first term of Eq.~(\ref{gammaffm}) came from $k=0$ mode 
of the sum, and it is nothing 
but the leading term of high  temperature expansion,
$\beta \rightarrow 0$. 
The difference of two divergent terms in the square brackets
is finite, however, we are unable to give the result in a closed form.
Thus, we restrict consideration to 
leading terms of high temperature expansion
and understand the basic thermal form factor as,
\begin{equation}
	\gamma^{\beta} ({-\triangle})= 
	\frac{ 4   \pi    }{  \beta   \sqrt{   - \triangle}    } \
	{\mathrm arctan}
	\Big(
      	\frac{  \sqrt{ -  \triangle } }{ 2 m  }\Big)+ {\mathrm O}[\beta],
	 \  \  \beta \rightarrow 0.     \label{arctan0}
\end{equation}
The main nonlocality is contained  in the leading term (\ref{arctan0}).
Subleading terms combined with vacuum contributions
are not important at high temperatures.
The full table of form factors $\gamma_i (q) $,\  $i=1, ... 4$, 
in terms of $ q =  \frac{m}{\sqrt{-\triangle}} $ reads
\begin{eqnarray}
	\gamma_1 (q)  &=& 
	\frac{ \pi q}{\beta m} 
	\left[ -\frac{5}{12} q - q^3 + (1+ 4 q^2)^2 \,
	{\mathrm arctan}
	\Big(
      	\frac{1}{ 2 q }     
	\Big) \right]                    \label{Garctan1}
\\ 	\gamma_2(q)&=& 
	 \frac{ \pi q}{24 \beta m} 
	\left[ \frac{13}{14} q - 3 q^3 -  
	\Big(\frac{11}{3}+ 28 q^2 + 48 q^4 \Big) \
	{\mathrm arctan}
	\Big(
      	\frac{1}{ 2 q }     
	\Big) \right]               \label{Garctan2}
\\	\gamma_3( q )&=& 
	 \frac{ \pi q}{ 2 \beta m} 
	\left[ 2 q +    	(1 -  4 q^2) \
	{\mathrm arctan}
	\Big(
      	\frac{1}{ 2 q }     
	\Big) \right]                     \label{Garctan3}
\\	\gamma_4(q )&=& 
	\frac{ 2 \pi q}{\beta m} \
	{\mathrm arctan}
	\Big(
      	\frac{1}{ 2 q }     
	\Big).  \label{Garctan4}
\end{eqnarray}     
                                     
For practical purposes of physical applications
we need to know  spectral form representations 
for (\ref{Garctan1})--(\ref{Garctan4}).
A spectral form for the basic form factor (\ref{arctan0}) is obvious,
\begin{equation}
	\gamma^{\beta} ({-\triangle})= 
	\frac{ 8   \pi }{  \beta} \,
 	\int_{m}^{\infty} \! {\mathrm d} \tilde{m} \,
      	\frac{1}{ 4 {\tilde{m}}^2 - \triangle }.       \label{spectrM}
\end{equation}
Its massless limit immediately gives (\ref{spectral}).
Spectral forms for form factors with subtractions
are obtained similarly (see appendix~\ref{C}).   
Then, all form factors  (\ref{Garctan1})--(\ref{Garctan4})
admit  the form,
\begin{equation}
	\gamma_i^{\beta}({-\triangle})= 
	\frac{ \pi }{ \beta} \,
 	\int_{m}^{\infty} \! {\mathrm d} \tilde{m} \,
  	\rho_i (\tilde{m}^2) \,
  	 \frac{1}{ 4 {\tilde{m}}^2 - \triangle },      \label{massspectrT}
\end{equation}
where mass spectral weights are given in the table,
\begin{eqnarray}
	\rho_1 ( \tilde{m}^2)&=& 
	\frac{ 1 }{ 4 } \,
  	\Big(1- \frac{2 m^2}{{\tilde{m}}^2}  + 
	\frac{m^4}{{\tilde{m}}^4}\Big), 
\\ 
	\rho_2 ( \tilde{m}^2)&=&  \mbox{}
	-\frac{1}{ 32 } \,
  	\Big( \frac{25}{9} - \frac{14}{3}  \frac{m^2}{{\tilde{m}}^2}  + 
	\frac{m^4}{{\tilde{m}}^4}\Big),
\\
	\rho_3 ( \tilde{m}^2)&=& 
      	\mbox{} - \frac{1}{3} +  \frac{m^2}{{\tilde{m}}^2},
\\    
	\rho_4 ( \tilde{m}^2)&=&  4.
\end{eqnarray}

Now we need to complete our derivation with the regularized  
free energy at zero temperature $F^{\infty}$.
Nonlocal effective action for massive fields  in an arbitrary 
spacetime dimension has been calculated first  by 
Avramidi \cite{Avram-PLB84}. 
His approach is a direct summation of 
derivatives in a massive field theory,
but we can make use of  the massless heat kernel (\ref{TrK}) 
obtained with the covariant perturbation theory and 
arrive at the same result. 
We compute zeta function regularized   effective action  
according to the equation,
\begin{equation} 
	{W}^{\infty}_{\mathrm{ren}}= 
	\mbox{}-\frac12 \frac{\partial}{\partial\epsilon} 
	\left[ 
	\frac{\mu^{2\epsilon}}{\Gamma(\epsilon)} 
	\int_0^{\infty}\! \frac{{\mathrm{d} }s}{s^{1-\epsilon}}\,  
	{\mathrm e}^{-s m^2}
	{\mathrm{Tr}} {K}(s)  
	\right]_{\epsilon=0}.         \label{masszeta-reg} 
\end{equation} 
Then, we get the following result for zero temperature free energy
(the specific form of the effective action in Ref.~\cite{Avram-PLB84}  
in four dimensions),
\begin{eqnarray}
          F^{\infty}_{\mathrm ren} &=& \mbox{} - \frac{1}{32 \pi^2}
	\int \! {\mathrm d}^3 x\, g^{1/2}  
	{\mathrm tr}\left\{
             - \frac{m^4}{2} 
\left(  \ln  \Big(  \frac{m^2}{\mu^2} \Big) - \frac{3}{2}  \right) \hat{1}
	\right. \nonumber \\ && \mbox{}   
	 + m^2 \left(  \ln\Big( \frac{m^2}{\mu^2} \Big) -1  \right) \hat{P}	
        	+\Big[ R_{i j } \gamma_1(-\triangle) R^{i j }
	+ R \gamma_2(-\triangle) R 
	\nonumber\\ && \left. \mbox{}
	+ \hat{P} \gamma_3(-\triangle) R
	+ \hat{P} \gamma_4(-\triangle) \hat{P}   
              \Big]    	
	+{\mathrm{O}}[\Re^3] 	 \label{efacM} \right\},
\end{eqnarray}                                   
where form factors $\gamma_i$ are given in terms of
dimensionless argument $ q =  \frac{m}{\sqrt{-\triangle}} $    
by the following expressions,
\begin{eqnarray}	
	\gamma_1 (q) &=& 
	\frac1{60} \left[   
	- {\mathrm ln}\Big(  \frac{m^2}{\mu^2}\Big)
	+ \frac{46}{15}
	+  \frac{56}{3} q^2
	+ 32 q^4 \right.
\nonumber\\ && \left. \mbox{}  
	- 2 \Big(  1 + 4 q^2  \Big)^{5/2}
	{\mathrm arctanh} \left(  \frac{1}{\sqrt{1+4 q^2 }}  \right)
	\right], \label{gamma1M} 
\\  [2mm]  
	\gamma_2 ( q )& =&  
	\frac1{180}   \left[
	{\mathrm ln} \Big( \frac{ m^2 }{ \mu^2 }\Big)-\frac{97}{30}
	-  17 q^2  - 12  q^4 
	\right.\nonumber\\ && \left. \mbox{}  
	+  2 \sqrt{1+ 4 q^2 } \Big(  1 +  8  q^2 +  6   q^4 \Big)
	{\mathrm arctanh} \left(  \frac{1}{\sqrt{1+ 4 q^2 }}  \right)
	\right],  \label{gamma2M} 	  
\\ [2mm]   
	\gamma_3 ( q )& =&  
	\frac1{6}   \left[
	- \frac13 -  4 q^2  
	+ 4  q^2  \sqrt{1+ 4 q^2  }\,\, 
	{\mathrm arctanh} \left(  \frac{1}{\sqrt{1+4 q^2 }}  \right)
	\right], \label{gamma3M}
	\\ [2mm]  
	\gamma_4 ( q )& =&  
	\mbox{} - \frac12 {\mathrm ln} \Big( \frac{ m^2 }{ \mu^2 }\Big) +1  
	-  \sqrt{1+ 4 q^2  }\,\, 
	{\mathrm arctanh} \left(  \frac{1}{\sqrt{1 + 4 q^2 }}  \right). 
\label{gamma4M} 	 	
\end{eqnarray}  
This effective action may look more similar to Eq.~(\ref{efacz})
if the inverse  hyperbolic tangents  in  functions $\gamma_i$ 
are expressed in terms of logarithms.
We have to remark here that form factor $\gamma_3$  
is different from the others.
Similarly to that  of massless fields  it  
does not depend on the  regularization 
parameter $\mu$.
However, Eq.~(\ref{gamma3M})
is nonlocal in contrast to local (\ref{gamma3}).  
Of course, in the zero mass limit  
Eqs.~(\ref{efacM})--(\ref{gamma4M})
turn to Eqs.~(\ref{efacz})--(\ref{gamma4}).

Finally, it should be noted that in order  to use Eq.~(\ref{efacM}) for
specific physical models, the convenient  way to work with form factors 
(\ref{gamma1M})--(\ref{gamma4M}) is 
to treat them in a spectral form representation \cite{Schwin-book89}.
Then, the following mass spectrum integral is to be used,
\begin{equation}
\frac{4}{\sqrt{-\triangle}} \frac{1}{\sqrt{4 m^2 - \triangle }}
{\mathrm arctanh} 
\left(  \frac{\sqrt{-\triangle }}{\sqrt{ 4 m^2 - \triangle}}  \right)
= \int_m^{\infty}\! {\mathrm d} \tilde{m} \,
\frac{1}{\sqrt{   {\tilde{m}}^2 - m^2}} \,
\frac{1}{4 {\tilde{m}}^2 - \triangle}.       
\end{equation}


               \section{Free energy of  massless   spinor   fields}
\label{spinor}                              

               
In this section we consider the massless 
Dirac spinors  $\psi$ in a Euclidean ultrastatic 
spacetime at finite temperature. 
The massless covariant Dirac equation is taken as
  	\begin{equation}
	\not\hspace{-1mm}\nabla \psi=0,\label{Dirac}
	\end{equation}  
where the standard notation
$ \not\hspace{-1mm}\nabla =\gamma^\mu\nabla_\mu$
is used (see \cite{FrolFurs-hepth9802} for general definitions). 
The method of calculation of the effective action
for spin-$1/2$ fields,  $W_{(1/2)}$ ,
is similar to the one for spin-$0$ fields. 
The main difference is that fermions are antiperiodic in 
the Euclidean time and,
therefore, they  satisfy boundary conditions (\ref{fields})  
with the minus sign.
The local form of the one-loop effective action 
$W_{(1/2)}=- {\mathrm Tr} \ln \not\hspace{-1mm}\nabla$ 
was studied first in Ref.~\cite{DeWitt-book65}.
It is  defined in
terms of the heat kernel (or zeta function) of operator  (\ref{Dirac}),
however,  following  DeWitt's idea
we consider the  squared operator $\not\hspace{-1mm}\nabla^2$, thus,    
\begin{equation}
	W_{(1/2)}^{\beta}=\beta F_{(1/2)}^{\beta}=
\frac12 \int_0^{\infty}\! \frac{{\mathrm d} s}{s}\,  
	{\mathrm Tr}  K _{(1/2)}^{\beta}(s),          \label{spinorW}
\end{equation}                
with the heat kernel $K _{(1/2)}(s)$  corresponding to
the squared Dirac operator.  It can be shown \cite{DeWitt-book65} that
the heat kernel  (Green function) of 
the operator $\not\hspace{-1mm}\nabla^2$
is equivalent  to the spinor heat kernel which is a solution of the equation,
\begin{equation} 
	\left\{
	{{\mathrm d} \over{\mathrm{d}}s}
	- \hat{1} \Big[  \Box -{1\over 4}  R(x)  
	\Big]
	\right\}
	\hat{K}_{(1/2)}(s|x;x')=
\hat{1}\delta(s)\delta(x;x').               \label{spinorHK}
\end{equation}

One can represent the heat kernel (\ref{spinorHK})  at finite temperature 
in a form of the image sum 
\cite{Dowk-QG84,DowkSch-NPB89},
	\begin{equation}
    \hat{K}^\beta_{(1/2)}
    (s\mid\tau,\mbox{\boldmath $x$};\tau',\mbox{\boldmath $x'$}) =
	\sum_{n=-\infty}^\infty (-1)^n
	\hat{K}_{(1/2)} (s\mid\tau,\mbox{\boldmath $x$};
	\tau'+\beta n,\mbox{\boldmath $x'$})         \label{sum-spinor}
	\end{equation}                         
 (cf. Eq.(\ref{sumK})).
Because time dependence of the heat kernel 
in ultrastatic spacetimes factorizes out (\ref{splitK}), 
the  trace of the heat kernel can be written in the form
\begin{equation} 
	{\mathrm{Tr}}{K}^{\beta}_{(1/2)}(s)=  
	\theta_2\Big(0,{\mathrm{e}}^{-\frac{\beta^2}{4s}}  
	\Big)  \frac{\beta}{(4 \pi s)^{1/2}}
	\int {\mathrm d}^3 x \,   
	{\mathrm{tr}}\, {K}^{(3)}_{(1/2)} 
	(s|\mbox{\boldmath $x$}, \mbox{\boldmath $x$}),    \label{fermiFE}
\end{equation} 
where $\theta_2$ is the Jacobi theta function.
It is convenient 
to use the $\theta$-functions identity
\begin{equation}
    \theta_2(0,{\mathrm e}^{-z})=
    \theta_3(0,{\mathrm e}^{-z/4})
    -\theta_3(0,{\mathrm e}^{-z})
\end{equation} 
and express the heat kernel trace 
at finite temperature (\ref{fermiFE}) in terms of 
the Jacobi  functions $\theta_3$.   
Thanks to this fact \cite{DowkSch-NPB89,Dowk-QG84} 
 we do not have to repeat all calculations and 
can  get the free energy of Dirac spinors at finite temperature  
using mathematical derivations of section~\ref{massless}. 
Spinor  thermal form factors $\gamma_{(1/2)}$
are obtained then
by a simple combination of scalar form factors $\gamma$
\begin{equation}       
	 \gamma_{(1/2)} \big( \beta \sqrt{-\triangle} \big) 
= - \left[ 2 \gamma \big( 2 \beta \sqrt{-\triangle} \big) 
- \gamma \big(  \beta \sqrt{-\triangle} \big)  \right].
\end{equation}

To form the operator of  Eq.~(\ref{spinorHK}) the potential term 
 should be taken  $\hat{P}=-\frac{1}{12} R \hat{1}$.
The commutator curvature in  not  zero 
when covariant derivatives act  on   spinors, 
but we need to know only that 
${\mathrm tr} \hat{\cal R}_{\mu\nu} \hat{\cal R}^{\mu\nu} =
-\frac18 R_{\mu\nu\alpha\beta} 
R^{\mu\nu\alpha\beta}  {\mathrm tr} \hat{1}$,
where the squared Riemann  tensor must be expressed 
via Ricci tensor and scalar curvature
with help of the Gauss-Bonnet identity \cite{BGVZ-JMP94-basis}.
All other matrix structures are reduced to ${\mathrm tr} \hat{1}=4$. 
Taking into account these properties
the free energy of  massless spinors reads   
(from now on we omit spinor indices {\tiny (1/2)})
\begin{eqnarray}
	 {F}^{\beta}_{\mathrm{ren}} -  
         {F}^{\infty}_{\mathrm{ren}}&=&
	\mbox{}- 
	\int \!  {\mathrm{d} }^3 x \,  {g}^{1/2}\, 
        \Big\{
	\frac{7 \pi^2}{180 \beta^4} -
	 \frac{1}{144 \beta^2} R
\nonumber\\ && \mbox{}
	+  \frac1{8\pi^2} \Big[
	R_{i j} \gamma^{\beta}_{1}(-\triangle) R^{i j} 
	+ R \gamma^{\beta}_{2}(-\triangle) R \, \Big]
	        +{\mathrm O}[R^3]\Big\},  \label{efacTspinor}
\end{eqnarray}                                              
where the thermal spinor form factors
\begin{equation}       
	 \gamma_{i} ( \beta \sqrt{-\triangle}) = 
    \int_0^\infty\!\! {\mathrm d}t \,\, g_{i} (t)
	\left[ 
	\frac{2\pi}{ \beta  {\sqrt{ -\triangle}}\,  t}\, \frac{1}{{\mathrm sh}
	(2\pi t/(\beta{\sqrt{ -\triangle}}))} 
	- \frac{1}{t^2}   \right],   \label{gammaTspinor}
\end{equation}
with the trigonometric  polynomials
\begin{eqnarray}       
	 g_1 (t)& =&  
	 \mbox{} -\frac14 
    \Big(  \frac{\cos (t)}{t}- 3 \frac{\sin (t)}{t^2} \Big)
	+ \frac32  \Big(  \frac{\cos (t)}{ t^3}-\frac{\sin (t)}{ t^4}\Big), \\   
	 g_2 (t)&=&   \mbox{}
    - \frac1{16} \left[
	  2  \frac{\sin (t)}{t^2} -  \frac{\cos (t)}{t} 
	+ 3\Big[ {\cos (t)\over t^3}-{\sin(t)\over t^4}\Big] 
\right].  
\end{eqnarray} 
Note, that the only  difference of 
(\ref{gammaTspinor}) from  (\ref{gammaTgen})
is the hyperbolic sinus  instead of  the hyperbolic tangent.

This  result is to be combined with 
the regularized  contribution  $F^{\infty}_{\mathrm ren}$
of the image sum (\ref{sum-spinor}).
The  zeta regularized effective action (\ref{spinorW}) 
at zero temperature takes the  form,
\begin{equation}  
         {F}^{\infty}_{\mathrm{ren}} = 
	-  \frac1{8\pi^2}  \int\! 
	{\mathrm d}^3 x \,  {g}^{1/2}   
        {\mathrm tr}\Big\{ R_{i j } \gamma_1(-\triangle) R^{i j }
	+ R \gamma_2(-\triangle) R 
             +{\mathrm{O}}[\Re^3]\Big\}.       \label{efacz-spinor}
\end{equation}
where zero temperature spinor form factors 
$\gamma_i(-\triangle)$,\ $i=1, ... 2$, are
\begin{eqnarray}	
	\gamma_1(-\triangle) &= & \mbox{}
	-\frac1{40} \left[ {\ln}\Big(-\frac{\triangle}{\mu^2}\Big)
	-\frac{12}{5}\right], 
\\ 
	\gamma_2 (-\triangle) & = & \frac1{120}   \left[
	{\ln}\Big(-\frac{\triangle}{\mu^2}\Big)
    -\frac{77}{30}\right].   
\end{eqnarray}  
The main result for the finite  temperature free energy
$F_{\mathrm ren}^{\beta}$ is the sum of 
Eqs.~(\ref{efacTspinor}) and (\ref{efacz-spinor}).
It is an essentially nonlocal functional.
On the other hand it is well known that only local terms
survive in high temperature limit \cite{Dowk-QG84}.
To check this property we consider the high temperature limit 
 of our result.

The calculation of  $\beta \rightarrow 0$ asymptotic of 
the thermal form factors (\ref{gammaTspinor}) are analogous
to similar derivations in section~\ref{massless}. 
A different hyperbolic function appearing in thermal form
factors, namely, the hyperbolic sinus, 
results in the absence of linear in temperature  nonlocal terms
in high temperature expansion, for example,
the basic form factor ($g_i (t)= - 2 \sin(t) $ in (\ref{gammaTspinor}))  
reads 
\begin{eqnarray}
{\gamma} ( \beta \sqrt{-\triangle}) &=&   
    -  2 \Big[   
{\mathrm C} +  \ln \Big(\frac{\beta \sqrt{-\triangle}}{\pi}\Big)  - 1   
    \Big] 
\nonumber\\&& \mbox{}
+  \frac{7 \zeta (3)}{24  \pi^2} \beta^2  ({- \triangle})
-  \frac{31 \zeta (5) }{640 \pi^4}  \beta^4  ({- \triangle})^2
+  {\mathrm{O}}[\beta^6],
\ \ \ \beta \rightarrow 0.                       
\end{eqnarray}                  
Uniting the vacuum free energy (\ref{efacz-spinor})
and the high temperature expansion of   (\ref{efacTspinor})
into one expression, we observe that the logarithmic nonlocality
also cancels in the sum.  The resulting 
$T=1/\beta \rightarrow \infty$ expansion
looks like
\begin{eqnarray}
	 {F}_{\mathrm{ren}}^{\beta}&=&
	\mbox{}-\int\!{{\mathrm d}}^3 x\,  {g}^{1/2}\, 
        	\left\{\frac{7 \pi^2}{180\beta^4}
	-\frac{1}{144\beta^2} R  \right.
\nonumber\\   && \mbox{}
	+\frac{1}{16\pi^2} \Big(  
	\ln (\frac{\beta\mu}{\pi})+{\mathrm{C}}\Big) \,
	\Big[
	\frac{1}{30}  R R 
	-\frac{1}{10} R_{ij} R^{ij}  \Big]  
\nonumber\\&&\mbox{}
	+\beta^2 \frac{7}{128}  \frac{\zeta (3)}{ \pi^4} 
	\Big[      
	\frac{1}{280} R \triangle  R
	- \frac{1}{84}  R_{ij}\triangle  R^{ij}
  \Big]                                 
\nonumber\\&&\mbox{}
	+\beta^4 \frac{93}{1024} \frac{\zeta (5)}{ \pi^6} 
	\Big[          
     \frac{1}{3780} R {\triangle}^2  R 
	- \frac{1}{1080}  R_{ij} {\triangle}^2   R^{ij}
 \Big] 
\nonumber\\&&\left. \mbox{}                 
	+{\mathrm{O}}[R^3]
	+{\mathrm{O}}[\beta^6] \vphantom{I}  \right\},
 	\  \  \beta \rightarrow 0.            \label{highTspinor}
        \end{eqnarray}                      
Combinations of quadratic in curvatures terms 
in the square brackets are the functional traces
of  Schwinger-DeWitt coefficient $a_2$, $a_3$, and $a_4$
after  substituting the Riemann curvature with the Ricci
tensor and scalar curvature
\cite{DeWitt-book65,BarVilk-PRep85,BGVZ-JMP94-asymp}.  
Expression (\ref{highTspinor}) coincides with the massless limit 
of a similar expression in Ref.~\cite{DowkSch-NPB89}, which corrects
apparent misprints in  higher $\beta$-order  
terms of Ref.~\cite{Dowk-QG84}.    

Massive spinor fields can be treated in a similar way.
At finite temperature the corresponding effective action is nonlocal.
We do not present the result  here because of its complexity. 
In the physically interesting limit of high temperatures
the effective action of fermi fields becomes local and can be found
in Ref.~\cite{DowkSch-NPB89}.
A small correction should be made to 
the result of \cite{DowkSch-NPB89}
because the effective actions corresponding to the first order
operator $(\not\hspace{-1mm}\nabla + m)$
and the squared one 
differ by local terms (see Eq.~(4.2) of Ref.~\cite{Deser}).


\section{Conclusions} 
\label{conclusion}

                                      
Free energy of quantum fields in ultrastatic (optical) spaces  
has been a subject of  study in a large  number of papers 
\cite{DowkKenn-JPA78,DowkCrit-PRD77,Dowk-QG84,%
Kirsten-JPA91,Kirsten-CQG93}.
When effects of gravitational fields are negligeable,  
the metric under consideration is flat, and  the spacetime 
is automatically ultrastatic. 
But even this simple case has not been studied sufficiently, 
as most works are concerned with  finding the effective action either  
on constant background fields or (and)  at very  high temperature 
\cite{Kapusta-book89,DolJack-PRD74,HabWeld-PRD82}. 
However, rapidly oscillating background fields  generate  
nonlocal terms in the effective action which 
contribute to  vacuum polarization effects 
\cite{FrolVilk-PLB81,Vilk-Gospel,MirzVilk-PRL95,MirzVilk-97}.   
Needless to say, that  an interesting and 
important high temperature limit 
of the effective action is still only an asymptotic,  
and knowledge of  the effective action behavior 
at arbitrary finite temperature is necessary. 
 
We have obtained nonlocal structures of the one-loop Euclidean
effective action and free energy for thermal fields in asymptotically
flat ultrastatic curved spacetimes. For nonconformal massless scalar
fields this expression has been  found at arbitrary finite temperatures. 
Explicit formulae for the high temperature limit 
of this general expression
have been obtained.
For massive scalar fields the free energy is derived 
in the high temperature limit. With help of these results
we calculated also the free energy of massless spinor fields.

The calculated one-loop Euclidean 
effective action is known to be a generating functional
of one-particle irreducible Green functions 
\cite{DeWitt-book65,Vasil'ev-book98}.
Therefore, variations of 
the nonlocal effective action over background fields 
generate the Green functions
\cite{gen-functonal}
and the energy-momentum  tensor \cite{MirzVilk-PRL95},
while variations of the free energy over thermodynamical variables, 
such as  temperature, provides one with thermodynamical potentials, 
entropy, etc.,  \cite{DowkKenn-JPA78,FrolFurs-hepth9802}.           

We would like to emphasize again that the technique
we adopted, namely, 
the nonlocal covariant perturbation theory \cite{CPT2},
is a  limit  opposite to the effective potential method 
\cite{DolJack-PRD74,Avram-JMP95}
and  the  derivative expansions 
\cite{MossTomsWright-PRD92,Kirsten-CQG93,HuCritStyl-PRD87}
suitable for constant or slowly changing background fields.
Therefore, to make comparisons one has to expand
our  nonlocal results in powers of 
the derivatives and compare them with
the corresponding terms  in the known derivative expansions
expanded, in turn, in powers of field strengths. 
Our results are in agreement with all known expansions of this kind.

It is important to stress the value of 
the nonlocal effective action approach.   
As has been  shown in section~\ref{massless} the effective action
for massless fields is naturally infrared finite,
which  demonstrates that infrared infinities
are absent both at zero and finite temperatures.
As for the applications of the covariant perturbation
theory to other models, one can refer to 
Refs.~\cite{OstrVilk-JMP88,BGVZ-NPB95},
where  spinor quantum electrodynamics 
in flat spacetime and scalar electrodynamics with gravity 
in the context of conformal anomaly have been studied.

We used in this paper  the imaginary time formalism,
which means that the notion of the global periodic time is introduced.
This greatly reduces the class of physical systems and spacetimes
that can be considered.
Alternatively,  there exists  real time formalism 
\cite{Keldysh-JETP65} 
which can give a covariant form of the partition function 
\cite{GribDonoghHolst-AP89} and free energy.
To work out  nonlocal structures of the effective action 
using the perturbation theory, which is 
applicable to nonequilibrium systems and
covariant from the outset, 
one should derive it from the scratch. 
It would be an extremely interesting project to do.

Obvious next step is to apply conformal transformation
technique \cite{DowkSch-PRD88,DowkSch-NPB89} 
to the obtained results and find the nonlocal
free energy on a more general static gravitational backgrounds.
This problem is facilitated by the fact that 
conformal transformations of  the nonlocal effective action
are  studied in detail in Refs.~\cite{MirzVilkZhyt-95}.   
In order to  study particle creation by external gravitational fields 
the third curvature order is required \cite{Vilk-CQG92,MirzVilk-97}. 
Our intention here was  to develop a  general method   
for  calculations of nonlocal free energy, which is the most transparent
while working in the second  order in curvatures.
Derivation of the next perturbative order, if needed,  will not pose
serious technical problems.  

Effects of space boundaries are also interesting and important 
\cite{DowkKenn-JPA78,Kirsten-JPA91}
and they deserve investigation, 
however, to do so by means of curvature expansion
a major revision  of covariant perturbation theory
is required.    

The results of this paper can be generalized to the very important 
case of background gauge fields.
Thus requires more  caution in dealing with
temporal components of gauge fields
and leads to the appearance of other noncovariant terms
in the effective action \cite{LeonZeln-PLB92}. 
It is not surprising, since at finite temperatures Lorenz 
invariance is broken. But the general structure of 
nonlocalities in the effective action  is still the same. 
                                                                                                                                 

\section*{Acknowledgments}


We thank V. Frolov, D. Fursaev, and R. Kobes for fruitful  discussions.
This work was partially supported by
National Science and Engineering Research Council
of Canada. Yu. G. is supported also
by CITA National Research Fellowship, 
and A. Z. is grateful to the Killam Trust for its financial support.


			\appendix


	\section{Massless form factors with subtractions}

\label{B}


Complex form factors entering the trace of the heat kernel (\ref{TrK}) 
are to be treated 
similarly to the basic one, as is described in section \ref{massless}.
The heat kernel form factor with one subtraction 
generates a thermal form factor in the effective action by the following
computational procedure,  
\begin{equation}  
	\eta_{1}  (\beta \sqrt{-\triangle})
	=
	\int_0^\infty { {\mathrm{d}} s\over s}
	\left[\theta_3\Big(0,{\mathrm e}^{-{\beta^2\over 4s}}\Big)-1\right]
	\frac{f(-s\triangle)-1}{s\triangle}.
\end{equation}  
Using integral representation,
\begin{equation}
	\frac{ f (- s \triangle )- 1 }{ s \triangle }=
	\int_0^1\!\! {\mathrm{d}}\alpha_1 \alpha_1 (1-\alpha_1) \!
	\int_0^1\!\!  {\mathrm{d}} \alpha_2  \,
	{\mathrm e}^{\alpha_1(1-\alpha_1)\alpha_2 s \triangle}, \label{1sub}
\end{equation}
and  taking the integral over proper time, 
we reduce it  to the form,  
\begin{equation}
	\eta_{1} (z) =
	4 \int_0^1\!\! {\mathrm{d}}\alpha_1 \, \alpha_1 (1-\alpha_1) \!
	\int_0^1\!\!  {\mathrm{d}} \alpha_2  \,
	K_0 ( n z \sqrt{\alpha_2} \sqrt{\alpha_1 (1 - \alpha_1)} ), 
\end{equation}
where  $z=\beta \sqrt{-\triangle}$ is a new dimensionless variable.                                                                                   
This is a  table integral \cite{GradRyzhik} with respect to $\alpha_2$,
\begin{equation}
	\int_0^1 {\mathrm d} y\, y^2 \, K_0 ( b y ) = 
	-\frac1b K_1 (b) + \frac{2}{b^2},  \label{intK0}
\end{equation}
which, after  introducing a new variable, 
$y=2 \sqrt{\alpha(1-\alpha)}$, give us 
 \begin{equation}
	\eta_{1} (z) =
	- 4 \sum_{n=1}^\infty \left[ {1\over nz} 
	\int_0^1  {\mathrm d} y \, 
    {y^2\over\sqrt{1-y^2}}\, K_1(\frac{nzy}{2}) 
	- 2 {1\over n^2 z^2} \right].      \label{bessel1}
\end{equation}
Applying the Bessel function relationship,
\[
	K_1(y)= 
    \mbox{}-  { {\mathrm{d}}  \over {\mathrm{d}} y} K_0 (y),
\] 
to $K_1$   and using  (\ref{sinus}) for the resulting integral,  
we get,
 \begin{equation}
	\eta_{1} (z)=
	- 4 \sum_{n=1}^\infty \left[
	\int_0^\infty  {\mathrm{d}} t \, {\sin(t)\over (t^2+n^2 z^2/4)^2} 
	- 2 {1\over n^2 z^2} \right]. 
\end{equation}  
After integration by parts, to reduce the power 
of the denominator, $\eta_{1}$
admits the form similar to Eq.~(\ref{sinus}),
\begin{equation}
	\eta_{1}(z)=
	2 \sum_{n=1}^\infty 
	\int_0^\infty\!   {\mathrm{d}}  t\,
	\left(   {\sin(t)\over t^2} - {\cos(t)\over t}    \right)
	{1\over (t^2+n^2 z^2/4)}.
\end{equation}
With use of (\ref{mainsum})  it  gives the final answer,
\begin{equation}
	\eta_{1} (z)= \int_0^\infty \!\!  {\mathrm{d}} t 
	\left(\frac{\sin(t)}{t^2}-\frac{\cos(t)}{t}\right)
	\left[
	\frac{2\pi}{zt}\,    \frac{1}{{\rm th}(2\pi t/z)}   -\frac{1}{t^2}
	\right].            \label{hgamma1} 
\end{equation}

Let us treat now the form factor with two subtractions,
\begin{equation}
	\eta_{2} ( \beta \sqrt{-\triangle} ) = 
	\int_0^\infty  {{\mathrm{d}} s\over  s}
	\left[\theta_3\Big(0,{\mathrm e}^{-{\beta^2\over 4s}}\Big)-1 \right]
	\frac{f(-s\triangle)- 1- \frac16 s\triangle }{(s\triangle)^2}.
\end{equation} 
In the representation,
\begin{equation}
	\frac{f(-s\triangle)-1-\frac16 s\triangle}{s\triangle}=
	\int_0^1\!\!  {\mathrm{d}} \alpha_1\, {\alpha_1}^2(1-\alpha_1)^2 
	\int_0^1\!\!  {\mathrm{d}} \alpha_2\, \alpha_2 
	\int_0^1\!\!  {\mathrm{d}}  \alpha_3 \, 
{\mathrm e}^{\alpha_1(1-\alpha_1)\alpha_2 \alpha_3 s\triangle}, \label{2sub}
\end{equation} 
it can be integrated first over the proper time.   
Integration over $\alpha_3$ is exactly Eq.~(\ref{intK0}), 
and integral over $\alpha_2$-parameter is
\begin{equation}
	\int_0^1 {\mathrm d} y\, y^2 K_1 ( b y ) = 
	-\frac1b K_2 (b) + \frac{2}{b^3}.  \label{intK1}
\end{equation}
Then, $\eta_{2}$  looks in variables $y$ and $z$ as
\begin{eqnarray}
	\eta_{2} (z)&=& 
	4  \sum_{n=1}^\infty \left[
	{1\over n^2 z^2} 
	\int_0^1 {\mathrm{d}} y\, {y^3\over\sqrt{1-y^2}}  \,
	K_2 (\frac{nzy}{2}) 
	+ {1\over 3} {1\over n^2 z^2}
	- 8 {1\over n^4 z^4 }. 
	\right]          \label{alphaK2}
\end{eqnarray}
It is possible to reduce (\ref{alphaK2}) to $\gamma$ and $\eta_{1}$
types of integrals employing the relation,
\begin{equation}
 K_2 (y) = \frac{2}{y} K_1 (y) + K_0 (y),
\end{equation}
\begin{eqnarray}
\eta_{2}(z)&=&
4\sum_{n=1}^\infty 
\left[
{1\over n^2z^2} 
\int_0^\infty \! {\mathrm d} t \, \sin(t)
\left(
{1\over t^2+n^2 z^2/4}
+6{1\over (t^2+n^2 z^2/4)^2}
\right)\right.  
\nonumber\\&&\left.\mbox{}
+ {1\over 3} {1\over n^2 z^2}
- 12 {1\over n^4 z^4 } \right].
\end{eqnarray}
We already know how to deal with $t$-integral, what is new here that is 
a sum,
\begin{equation}
\sum_{n=1}^{\infty} \frac{1}{n^2 z^2}\,\frac{1}{(t^2+n^2 z^2 /4)}=
\frac{\pi^2}{6 z^2 t^2}+ \frac{1}{8 t^4}-
\frac{\pi}{4 z t^3}\,\frac{1}{{\rm th}(2\pi t/z)}.
\end{equation}
which brings $\eta_{2}$ to the form,
\begin{eqnarray}
\eta_{2}(z)&=& 
\int_0^\infty \!\!   {\mathrm{d}} 
t \left( \sin (t) 
+ 3\Big[{\cos (t)\over t}-{\sin(t)\over t^2}
\Big]\right)
\left[\frac{2\pi^2}{3 z^2 t^2} 
-\frac{\pi}{ z t^3}\, \frac{1}{{\rm th}(2\pi t/z)} 
+ \frac{1}{2 t^4}
\right]
\nonumber\\&&\mbox{}
+\frac{2\pi^2}{9 z^2}.
\end{eqnarray}
And the final result reads,
\begin{equation}
\eta_{2}(z)=
\mbox{}-\frac12
\int_0^\infty\!\!  {\mathrm{d}}  t 
\left( \frac{\sin (t)}{t^2} 
+ 3\Big[{\cos (t)\over t^3}-{\sin(t)\over t^4}
\Big]\right)
\left[ 
\frac{2 \pi}{ z t}\, \frac{1}{ {\mathrm th}( 2 \pi t/z )} 
- \frac{1}{t^2}
\right].       \label{hgamma2} 
\end{equation}   


		\section{Massive form factors with subtractions}
 \label{C}

During the computation of remaining two form factors,
with one subtraction
\begin{equation}
\eta_{1}(\beta \sqrt{-\triangle}) = 
\frac{2 \sqrt{\pi }}{\beta }
\int_0^{\infty}  \frac{ {\mathrm{d}}  s}{ s^{3/2} } \,  
{ \mathrm{e}}^{ - s m^2 } \,
\frac{f(-s\triangle)-1}{s\triangle}, \  \beta \rightarrow 0, \label{m1}
\end{equation}                               
and with two subtractions                    
\begin{equation}
\eta_{2}(\beta \sqrt{-\triangle}) =
\frac{2 \sqrt{\pi }}{\beta}
\int_0^{\infty}  \frac{ {\mathrm d}  s}{ s^{3/2} } \,  
{ \mathrm e}^{ - s m^2 } \,
\frac{f (-s\triangle)- 1- \frac16 s\triangle }{(s\triangle)^2}, \label{m2}
\  \beta \rightarrow 0,
\end{equation}
we get rid of the image sum from the outset and keep only $k=0$ term. 

To perform the proper time integration
we need two singular integrals  that
are regularized by cutoff at the lower limit,
\begin{eqnarray}
A_1 (b)  \equiv 
\int_{\epsilon}^{\infty}  \frac{{\mathrm{d}}  s}{ s^{3/2} } \, 
 {\mathrm{e}}^{ - s b } & =&  
 - 2  \sqrt{\pi b } + \frac2{  \sqrt{\epsilon} },  \\
 A_2(b) \equiv
\int_{\epsilon}^{\infty}  \frac{ {\mathrm{d}}  s}{ s^{5/2} } \, 
{\mathrm{e}}^{ - s b } & =&  
\frac43 b^{3/2} \sqrt{\pi} +  \frac2{  {3 \epsilon}^{3/2} }
- \frac{ 2 b }{ \sqrt{ \epsilon } },
\end{eqnarray}
where $ b > 0 $.
The form factors (\ref{m1}) and (\ref{m2}) take then the form,
\begin{equation}
\eta_{1}(\beta \sqrt{-\triangle})= 
\frac{\sqrt{4 \pi }}{\beta  \triangle}\,  
\Big(
\int_0^1 \! {\mathrm{d}} \alpha   \, 
A_1(b(\alpha)) - A_1(m^2)  \Big),           \label{ma1}
\end{equation}
\begin{equation}
\eta_{2}(\beta  \sqrt{-\triangle})=
\frac{ 2 \sqrt{\pi} }{ \beta {\triangle}^2 }
 \Big(  \int_0^1 \! {\mathrm{d}} \alpha \, A_2 ( b(\alpha) )
 - A_2(m^2) - \frac1{6 \triangle} A_1(m^2) 
\Big),       \label{ma2}
\end{equation}                  
where $b(\alpha)= m^2 - \triangle \alpha (1 - \alpha)$.
Now we make use of regularized integrals  $A_1$ and $A_2$.
Functions $\eta_{1}$ and $\eta_{2}$ are regular,
therefore, all intermediate singularities 
in these equations  mutually cancel.
Integrations over $\alpha$-parameter do not pose a problem and they
are similar to the integral (\ref{alpha-int}). The only difference is that 
higher powers of integrand $b(\alpha)$  
bring extra factors at the arctangent function:
\begin{eqnarray}
	\eta_{1} (\beta \sqrt{-\triangle}) &=&
	\frac{\pi}{\beta  \sqrt{   - \triangle}   }  \left[
	\mbox{} - \frac{2 m}{  \sqrt{   - \triangle}   } +
	\Big( 1- \frac{4 m^2}{\triangle} \Big) \
	{\mathrm arctan}
	\Big(
      	\frac{  \sqrt{ -  \triangle } }{ 2 m }     
	\Big) \right],           \label{gM1}
\\
	\eta_{2} (\beta \sqrt{-\triangle})&=&
	\frac{\pi}{\beta  \sqrt{   - \triangle}   } \left[ 
  	 \mbox{} - \frac{5 m}{ 12  \sqrt{   - \triangle}   } 
	- \frac{ m^3}{ {(-\triangle)}^{3/2}}  \right.
\nonumber\\&&\mbox{}   \left.
    +	\Big( 1- \frac{4 m^2}{\triangle}  \Big)^2 \ 
	{\mathrm arctan}
	\Big(      	\frac{  \sqrt{ -  \triangle } }{ 2 m }     
	\Big)  \right].        \label{gM2}
\end{eqnarray}

Spectral forms for $\eta_{1}$ and  $\eta_{2}$ 
are readily obtained with help of basic spectral integral (\ref{spectrM})
applied to arctangent functions in (\ref{gM1})--(\ref{gM2}).
To transform operator factors to spectral weights we only need to know   
the identity,
\begin{equation}
\frac{1}{\triangle} 
\frac{1}{ 4 {\tilde{m}}^2 - \triangle } =
\frac{1}{4{\tilde{m}}^2 \triangle}
-  \frac{1}{4{\tilde{m}}^2}
\frac{1}{ 4 {\tilde{m}}^2 - \triangle}. 
\end{equation}
The spectral representations read,
\begin{equation}
	\eta_{1} (\beta \sqrt{-\triangle})= 
	\frac{ 2   \pi }{  \beta} \,
 	\int_{m}^{\infty} \! {\mathrm d} \tilde{m} \,
      	\Big(1- \frac{m^2}{{\tilde{m}}^2} \Big) \,
 	 \frac{1}{ 4 {\tilde{m}}^2 - \triangle }     
\end{equation}

\begin{equation}
	\eta_{2} (\beta \sqrt{-\triangle})= 
	\frac{ \pi }{ 4 \beta} \,
 	\int_{m}^{\infty} \! {\mathrm d} \tilde{m} \,
  	\Big(1- \frac{2 m^2}{{\tilde{m}}^2}  + 
    \frac{m^4}{{\tilde{m}}^4}\Big) \,
  	 \frac{1}{ 4 {\tilde{m}}^2 - \triangle }     
\end{equation}




\begin{thebibliography}{99}  
\bibitem{Matsubara59} 
	T. Matsubara, 
	Progr. Theor. Phys. {\bf 14} (1955) 351.
\bibitem{Fradk59} E. S. Fradkin, 
	Sov. Phys. Doklady, {\bf 4} (1959) 347;
	Nucl. Phys. {\bf 12} (1959) 465;
	Sov. Phys. JETP {\bf 36} (1959) 912.
\bibitem{MartSchwin-PR59} 
	P. C. Martin and J. Schwinger,
	Phys. Rev. {\bf 115} (1959) 1342.   
\bibitem{Kapusta-book89}
	J.  I. Kapusta,
	{\em Finite-temperature field theory}
	(Cambridge University Press, Cambridge, 1989).
\bibitem{DolJack-PRD74}
	L. Dolan and R. Jackiw,
	Phys. Rev. D {\bf 9} (1974) 3320.  
\bibitem{MossTomsWright-PRD92}
    I. Moss, D. Toms, and A. Wright,
    Phys. Rev. D {\bf 46} (1992) 1671.
\bibitem{Jackiw-PRD74}
	R. Jackiw, 
    Phys.Rev. D {\bf 9} (1974) 1686.   
\bibitem{Avram-JMP95}   
    I.  G. Avramidi, 
    J. Math. Phys. {\bf 36} (1995) 1557. 
\bibitem{DeWitt-book65}
	B. S. DeWitt,
	{\em Dynamical theory of groups and fields}
	(Gordon and Breach, New York, 1965).                  
\bibitem{BarVilk-PRep85}
        A. O. Barvinsky and G. A. Vilkovisky,
        Phys. Rep.  {\bf 119} (1985) 1.                                    
\bibitem{Ball-PRep89}
	R. D. Ball,
	Phys. Rep. {\bf 182} (1989) 1.
\bibitem{BraatPisar-PRD92} 
   E. Braaten and R. D. Pisarski, 
    Phys. Rev. D {\bf 45} (1992) 1827.     
\bibitem{DrumHorLandReb-PLB97}
	I. T. Drummond, P. R. Horgan, P. V. Landshoff, and A. Rebhan,
	Phys. Lett. B  {\bf 398} (1997) 326.     
\bibitem{Vilk-Gospel}
	G.  A. Vilkovisky,
	in {\em Quantum theory of gravity},
	ed. S. M. Christensen
	(Hilger, Bristol, 1984) p. 169.  
\bibitem{CPT1}
        A. O. Barvinsky and G. A. Vilkovisky,
        Nucl. Phys. B {\bf 282} (1987) 163.
\bibitem{CPT2}
        A. O. Barvinsky and G. A. Vilkovisky,
        Nucl. Phys. B {\bf 333} (1990) 471.
\bibitem{CPT3}
        A. O. Barvinsky and G. A. Vilkovisky,
        Nucl. Phys. B  {\bf 333} (1990) 512.
\bibitem{CPT4}
	A. O. Barvinsky, Yu. V. Gusev,
	G. A. Vilkovisky, and V. V. Zhytnikov,
	Report of the University of Manitoba (1993).                
\bibitem{DowkKenn-JPA78}
	J. S. Dowker and G. Kennedy,
	J. Phys.  A: Math. Gen. {\bf  11} (1978) 895.     
\bibitem{Furs-hepth9709}
        	D. V. Fursaev, 
	{\em Euclidean and canonical formulations of statistical
	mechanics in the presence of Killing horizons},
	preprint hep-th/9709213. 
\bibitem{FrolFurs-hepth9802}
    V.  P. Frolov and D.  V. Fursaev, 
	{\em Thermal fields, entropy, and black holes},
	preprint hep-th/9802010.        
\bibitem{FrolVilk-PLB81}
	V.  P. Frolov and G.  A. Vilkovisky,
	Phys. Lett. B {\bf 106} (1981) 307.        
\bibitem{Vilk-CQG92}
   G.  A. Vilkovisky, 
    Class. Quantum Grav. {\bf 9} (1992) 895.  
\bibitem{Allen-PRD86} 
	B. Allen,
	Phys. Rev. D {\bf 33} (1986) 3640. 
\bibitem{DowkCrit-PRD76}
    J.  S. Dowker and R. Critchley,
	Phys. Rev. D {\bf 13} (1976) 3224.
\bibitem{Elizal-book95} 
	E. Elizalde, 
	{\em Ten physical applications of
	spectral zeta functions}
	(Springer, Berlin, 1995).       
\bibitem{conform-trans}
	J.  S. Dowker,
	Phys. Rev.  D {\bf 33} (1986) 3150;
    M.  R. Brown and A. C. Ottewil,
    Phys. Rev. D {\bf 31} (1985) 2514;
    I. L. Bukhbinder, V. P. Gusynin, and P. I. Fomin,
    Sov. J. Nucl. Phys. {\bf 44} (1986) 534;
	J. S. Dowker,
	Phys. Rev. D  {\bf 39} (1989) 1235.
\bibitem{DowkSch-PRD88}
	J. S. Dowker and J. P. Schofield,
	Phys. Rev.  D {\bf 38} (1988) 3327.
\bibitem{DowkSch-NPB89}
	J. S. Dowker and J. P. Schofield,
	Nucl. Phys.  B {\bf 327} (1989) 267.   
\bibitem{DowkCrit-PRD77}
	J. S. Dowker and R. Critchley,
	Phys. Rev. D {\bf 15} (1977) 1484.        
\bibitem{BrownMacl-PR69}
    L. S. Brown and G. J. Maclay,
    Phys. Rev. {\bf 184} (1969) 184.  
\bibitem{GradRyzhik}
	I. S. Gradshteyn and I. M. Ryzhik,
	{\em Table of integrals, series, and products} 
	(Academic Press, New York, 1994).
 \bibitem{Gibbons-PLB77}
    G. W. Gibbons,
    Phys. Lett. A {\bf 60} (1977) 385.  
\bibitem{SDWcoeff}
	P. B. Gilkey,
	J. Diff. Geom. {\bf 10} (1975) 601;
    I. G. Avramidi, 
    Phys. Lett. B {\bf 238} (1990) 92. 
\bibitem{Schwin-PR51}
	J. Schwinger,
	Phys. Rev. {\bf 82} (1951) 664.            
\bibitem{GusZel-CQG98} 
	Yu. V. Gusev and A. I. Zelnikov, 
	Class. Quantum Grav. {\bf 15} (1998) L13.	  
\bibitem{BarFrolZel-PRD95}
	A. O. Barvinsky, V. P. Frolov and A. I. Zelnikov,
	Phys. Rev. D {\bf 51} (1995) 1741.             
\bibitem{BGVZ-JMP94-asymp}
	A. O. Barvinsky, Yu. V. Gusev,
	G. A. Vilkovisky, and V. V. Zhytnikov,
	J. Math. Phys. {\bf 35} (1994) 3543.  
\bibitem{Schwin-book89}
	J. S. Schwinger,
	{\em Particles, sources, and fields}
	(Addison-Wesley, Redwood, 1989) v. 2. 
\bibitem{HabWeld-PRD82}
	H. E. Haber and H. A. Weldon,
	Phys. Rev. D {\bf 25} (1982) 502.   
\bibitem{BranFrenk}
	F. T. Brandt and J. Frenkel, 
	Phys. Rev. D {\bf 55} (1997) 7808;
    F. T. Brandt and J. Frenkel,
	{\em The structure of the graviton self-energy 
	at finite temperature},
	preprint hep-th/9803155.
\bibitem{Avram-PLB84}
	I. G. Avramidi,
	Phys. Lett.  B {\bf 236} (1990) 443.  
\bibitem{Dowk-QG84}
	J. S. Dowker,
	in {\em Quantum theory of gravity},
	ed. S. M. Christensen
	(Hilger, Bristol, 1984) p. 103.                  
\bibitem{BGVZ-JMP94-basis}
	A. O. Barvinsky, Yu. V. Gusev,
	G. A. Vilkovisky, and V. V. Zhytnikov,
	J. Math. Phys. {\bf 35} (1994) 3525.  
\bibitem{Deser} 
	S. Deser, L. Griguolo and D. Seminara, 
    Phys. Rev. D {\bf 57} (1998) 7444.
\bibitem{Kirsten-JPA91}
    K. Kirsten,
    J. Phys. A: Math. Gen. {\bf 24} (1991) 3281.    
\bibitem{Kirsten-CQG93}
    K. Kirsten, 
    Class. Quantum Grav. {\bf 10} (1993) 1461.    
\bibitem{MirzVilk-PRL95}
	A. G. Mirzabekian and G. A. Vilkovisky,
	Phys. Rev. Lett. {\bf 75} (1995) 3974.
\bibitem{MirzVilk-97}
	A. G. Mirzabekian and G. A. Vilkovisky,
	Phys. Lett.  B {\bf 414} (1997) 123;
	A. G. Mirzabekian and G. A. Vilkovisky, 
    {\em Particle creation in the effective action method},
    preprint gr-qc/9803006.
\bibitem{Vasil'ev-book98}
	A. N. Vasil'ev,
	{\em Functional methods in quantum field theory
	and statistics}
	(Gordon and Breach, 1998).
\bibitem{gen-functonal}
	A. O. Barvinsky and G. A. Vilkovisky,
	in {\em Quantum Field Theory and Quantum Statistics}, vol. 1,
	eds. I. A. Batalin, C. J. Isham and G. A. Vilkovisky
	(Hilger, Bristol, 1987) p. 245;
	A. O. Barvinsky and Yu. V. Gusev,
	Class. Quantum Grav. {\bf 9} (1992)  383;
	A. O. Barvinsky and  Yu. V. Gusev,
	in {\em Heat Kernel Techniques and Quantum Gravity},
	ed. by S. A. Fulling
	(Texas A\&M University, College Station, Texas, 1995) p. 189.  
\bibitem{HuCritStyl-PRD87}
    B. L. Hu, R. Critchley, and A. Stylianopoulos,
    Phys. Rev. D {\bf 35} (1987) 510.      
\bibitem{OstrVilk-JMP88}
	A. A. Ostrovsky and G. A. Vilkovisky,
	J. Math. Phys. {\bf 29} (1988) 702.    
\bibitem{BGVZ-NPB95}
        A. O. Barvinsky, Yu. V. Gusev,
        G.  A. Vilkovisky, and V. V. Zhytnikov,
       Nucl. Phys. B {\bf 439} (1995) 561.      
\bibitem{Keldysh-JETP65}
	L. V. Keldysh,
	Sov. Phys. JETP {\bf 20} (1965) 1018. 
\bibitem{GribDonoghHolst-AP89}
    P. S. Griboski, J. F. Donoghue, and B. R. Hostein,
    Ann. Phys. (N.Y.) {\bf 190} (1989) 149.    
\bibitem{MirzVilkZhyt-95} 
	A. G. Mirzabekian, G. A. Vilkovisky, V. V. Zhytnikov, 
	Phys. Lett. B {\bf 369} (1996) 215;
	A.  O. Barvinsky, A. G. Mirzabekian, V. V. Zhytnikov,
	in {\em Quantum Gravity: 
	Proceedings of the Sixth Moscow 
	Quantum Gravity Seminar}, ed. V. Berezin
	(World Scientific, Singapour, 1997),  preprint gr-qc/9510037.                                                
\bibitem{LeonZeln-PLB92}
        A.  V. Leonidov and A.  I. Zelnikov,
        Phys. Lett. B {\bf 276} (1992) 122.    
\end{thebibliography}
\end{document}